\documentclass[a4paper,11pt]{article}
\usepackage{jheppub}
\usepackage{caption}
\usepackage{amsmath,amssymb,amsfonts}
\usepackage{blkarray}

\def\CA{{\cal A}}
\def\CB{{\cal B}}

\def\CD{{\cal D}}

\def\CF{{\cal F}}

\def\CH{{\cal H}}
\def\CI{{\cal I}}

\def\CK{{\cal K}}
\def\CL{{\cal L}}
\def\CM{{\cal M}}
\def\CN{{\cal N}}
\def\CO{{\cal O}}
\def\CP{{\cal P}}

\def\CS{{\cal S}}
\def\CT{{\cal T}}

\def\CV{{\cal V}}
\def\CW{{\cal W}}

\def\BI{{\mathbb I}}

\def\BN{{\mathbb N}}

\def\BR{{\mathbb R}}
\def\BS{{\mathbb S}}

\def\BV{{\mathbb V}}

\def\BZ{{\mathbb Z}}

\let\a\alpha
\let\b\beta

\let\d=\delta
\let\e=\epsilon
\let\z=\zeta
\let\h=\eta

\let\l=\lambda
\let\m=\mu
\let\n=\nu
\let\x=\xi
\let\p=\pi

\let\s=\sigma
\let\t=\tau 
\let\f=\phi
\let\c=\chi
\let\ps=\psi
\let\w=\omega

\let\G=\Gamma
\let\D=\Delta

\let\S=\Sigma

\def\vf{\varphi}

\let\pt=\partial
\let\hb=\hbar
\let\ra=\rightarrow
\let\lora=\longrightarrow

\def\SU{\mathrm{SU}}

\def\Li{\mathrm{Li}}

\def\Im{\mathrm{Im}}

\def\QD{\ps_\hb}

\newcommand{\paren}[1]{\left( {#1} \right)}

\def\pd{\partial}
\def\BV{{\rm BV}^{\BS(\vec{k})}}
\def\HF{\mathbf{HF}}
\def\dm{\cdot\vec{m}}
\title{\boldmath Non-unitary Haagerup-like TQFTs and RCFTs from generalized S-fold SCFTs}

\author{Kibok Jeong, Huijoon Sohn}
\affiliation{Department of Physics and Astronomy $\&$ Center for Theoretical Physics,\\
Seoul National University, 1 Gwanak-ro, Seoul 08826, Korea}

\emailAdd{boki0322@snu.ac.kr}
\emailAdd{hjson99@snu.ac.kr}

\abstract{Building on earlier work by Gang, Kim, and Lee, we extend the known class of 3-dimensional non-unitary topological quantum field theories (TQFTs) and 2-dimensional non-unitary rational conformal field theories (RCFTs). The TQFTs are obtained by applying topological twist to generalized S-fold superconformal field theories (SCFTs). They are related to the RCFTs through the bulk-boundary correspondence. For certain families, we propose simple lines in the TQFTs, together with expressions for the characters and modular $S$ matrices of their associated boundary RCFTs. For broader family, we propose modular $S$ and $T$ matrices for the TQFTs. For particular members of this family, the resulting modular matrices are identified with the generalized Haagerup-Izumi modular matrices up to Galois conjugation.}

\begin{document}
\maketitle
\flushbottom

\section{Introduction}
Recent developments in supersymmetric quantum field theory have provided physical realizations of 3-dimensional non-unitary topological quantum field theories (TQFTs)\cite{Closset:2026xjj,Gang:2021hrd,Jeong:2025xid,Gang:2024loa,Gang:2023rei,Kim:2025rog,Go:2025ixu}, many of which had previously been studied through axiomatic approaches\cite{Atiyah:1989vu,Reshetikhin:1990pr,Reshetikhin:1991tc,Turaev:1994xb}. 3-dimensional $\CN=4$ rank-0 superconformal field theories (SCFTs), characterized by trivial Coulomb and Higgs branches\cite{Gang:2018huc,Creutzig:2024ljv,Gang:2021hrd,ArabiArdehali:2024ysy,Creutzig:2026ajk,Baek:2024tuo}, can realize non-unitary TQFTs after topological twists\cite{Gukov:2020lqm,Garner:2022rwe,Creutzig:2021ext,Rozansky:1996bq}. This framework also enables a systematic study of non-unitary bulk-boundary correspondences\cite{Gang:2023rei,Gang:2024loa,Gang:2023ggt,Ferrari:2023fez,Dedushenko:2023cvd,ArabiArdehali:2024ysy,Baek:2024tuo,Gang:2025ykf,Witten:1988hf} between 3-dimensional TQFTs and 2-dimensional rational conformal field theories (RCFTs)\cite{Gang:2024loa,Gang:2023rei,Mathur:1988na,Zhu:1996gaq,Chandra:2018pjq,Mukhi:2019xjy,Duan:2022ltz,Duan:2022kxr,Gang:2025ykf}. BPS partition functions\cite{Kim:2009wb,Hama:2010av,Hama:2011ea,Benini:2015noa,Pestun:2016zxk,Dimofte:2017tpi,Gang:2009qdj} provide access to data of the non-unitary TQFTs and the associated boundary RCFTs.

Within this framework, the topological twists of a class of SCFTs called \textit{S-fold SCFTs}\cite{Gang:2022kpe,Gang:2023ggt,Assel:2018vtq,Garozzo:2018kra,Garozzo:2019ejm,Garozzo:2019hbf,Beratto:2020qyk,Arav:2021gra,Bobev:2021yya,Bobev:2023bxs,Imamura:2026dhk} were shown to realize a novel class of non-unitary TQFTs\cite{Gang:2022kpe}. Here, S-fold SCFTs are obtained by gauging the diagonal $\SU(2)$ subgroup of the global symmetry of the $T[\SU(2)]$ theory\cite{Gaiotto:2008ak} at a non-vanishing Chern-Simons level. Under the 3d-3d correspondence\cite{Terashima:2011qi,Closset:2026xjj,Choi:2020baw,Gang:2018wek,Baek:2025uev,Baek:2024tuo,Gang:2024tlp,Gang:2026iem,Choi:2022dju,Dimofte:2011py,Dimofte:2011ju,Gang:2025ykf}, the same theories are associated with once-punctured torus bundles($\S_{1,1}\times_{\vf}S^1$)\cite{Terashima:2011qi, Gang:2013sqa}, with certain family of monodromies $\vf$. These theories are $\CN=4$ rank-0 SCFTs, and their topological twists realize non-unitary TQFTs. The modular matrices of the resulting non-unitary TQFTs generalize Haagerup-Izumi modular matrices\cite{Asaeda_1999,Evans:2010yr,Evans:2015zga,Izumi:2000qa}, up to Galois conjugation. The bulk-boundary correspondence associates the resulting TQFTs with a novel class of non-unitary RCFTs\cite{Gang:2023ggt}.

\begin{figure}[h]
\centering
\includegraphics[width=.70\textwidth]{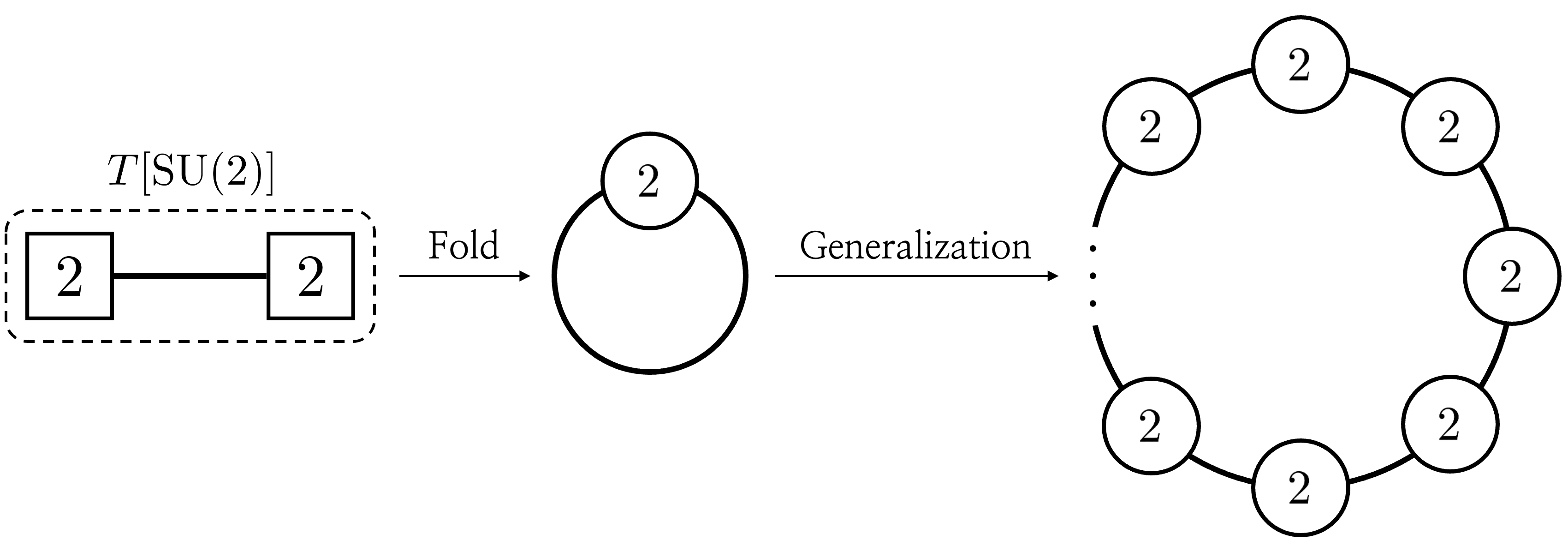}
\captionof{figure}{Schematic representation of a generalized S-fold SCFT. The $T[{\rm SU}(2)]$ theory is represented by a line segment.}
\label{fig: introfigure}
\end{figure}

These results suggest that generalizations of S-fold SCFTs may give rise to further families of non-unitary TQFTs and their associated RCFTs. Motivated by this possibility, in the present work we investigate a broader class of $\CN=4$ rank-0 SCFTs which we call \textit{generalized S-fold SCFTs} and study the non-unitary TQFTs and boundary RCFTs associated with them. The generalized S-fold SCFTs are constructed by gauging diagonal $\SU(2)$ subgroups of the global symmetries of multiple copies of the $T[\SU(2)]$ theory at non-vanishing Chern-Simons levels. From the perspective of the 3d-3d correspondence, these theories are associated with once-punctured torus bundles with more general monodromies. By the topological twists, the generalized S-fold SCFTs realize a broader class of non-unitary TQFTs.

Using BPS partition functions, we extract information about the resulting TQFTs and associated boundary RCFTs. For certain families of generalized S-fold SCFTs, we propose the sets of simple lines for the resulting non-unitary TQFTs. Using these simple lines, we derive expressions for the characters of their associated boundary RCFTs. We also propose modular $S$ matrices for these RCFTs. For a broader family of generalized S-fold SCFTs, we propose modular $S$ and $T$ matrices for the resulting non-unitary TQFTs. For particular theories, the modular $S$ and $T$ matrices agree, up to Galois conjugation, with those of the generalized Haagerup-Izumi modular data introduced in \cite{Evans:2010yr}.

The remaining part of this paper is organized as follows. In section \ref{sec: Theory labeled by 1pt torus}, we introduce the theories of interest and present complementary descriptions of them: as $\CN=2$ Chern-Simons-matter theories and as generalized S-fold SCFTs. In section~\ref{sec: abelian CSM theory}, we investigate certain families of theories using their abelian Chern-Simons-matter descriptions, focusing on the boundary RCFT data associated with the resulting non-unitary TQFTs. In section~\ref{sec: generalized S-fold}, we investigate a broader family of theories using their generalized S-fold SCFT descriptions, focusing on the modular matrices for the resulting non-unitary TQFTs. In the appendices, we provide computational details and consistency checks supporting the main results of section~\ref{sec: generalized S-fold}.

\section{Two descriptions of $T[\S_{1,1}\times_\vf S^1]$} \label{sec: Theory labeled by 1pt torus}
Given a 3-manifold $\CM_3$, a twisted compactification of the 6-dimensional $A_1$ $\CN=(2,0)$ theory on $\CM_3$ \textit{defines} a 3-dimensional $\CN=2$ theory $T[\CM_3]$. In this paper, we are interested in theories labeled by the once-punctured torus bundle associated with a monodromy $\vf\in{\rm SL}(2,\mathbb{Z})$\cite{Terashima:2011qi,Gang:2013sqa,Gang:2023ggt,Gang:2022kpe}
\begin{align}
    \CM_3=\S_{1,1}\times_\vf S^1\equiv \S_{1,1}\times I\bigr/\bigr( (x,0)\sim(\vf(x),1)\bigr)\;.
\end{align}
Here, we assume $|{\rm tr}\vf|>2$. This corresponds to the case in which $\S_{1,1}\times_\vf S^1$ is hyperbolic and the theory $T[\S_{1,1}\times_\vf S^1]$ becomes the 3-dimensional $\CN=4$ rank-0 SCFT\cite{Terashima:2011qi,Gang:2013sqa,Gang:2023ggt,Gang:2022kpe}.

We consider two UV descriptions of $T[\S_{1,1}\times_\vf S^1]$. The first one can be obtained from the Dimofte-Gaiotto-Gukov construction\cite{Dimofte:2011ju}. The DGG construction provides an $\CN=2$ abelian Chern-Simons-matter theory $T_{\rm DGG}[\vf]$ from the ideal triangulation of $\S_{1,1}\times_\vf S^1$. The $\CN=2$ supersymmetry is enhanced to $\CN=4$ along the RG flow and the IR fixed point is identified with $T[\S_{1,1}\times_\vf S^1]$, up to a decoupled unitary TQFT. We provide more details in section~\ref{subsec: TDGG}. The other one is the generalized S-fold SCFT $\BS(\vec{k})$. We can understand the twisted compactification as a two-step procedure\cite{Gang:2013sqa,Terashima:2011qi}. First, we compactify the 6-dimensional theory on $\S_{1,1}$. It yields the 4-dimensional $\CN=4$\footnote{We assume that the real mass parameter associated with ${\rm U}(1)_A$ is set to zero. This condition is essential for ensuring $\CN=4$ supersymmetry\cite{Terashima:2011qi}.} supersymmetric Yang-Mills theory with gauge group $G={\rm SU}(2)$, which has ${\rm SL}(2,\mathbb{Z})$ as its duality group. Next, we reduce the resulting theory on $S^1$ with a twist by $\vf$. Then we obtain $T[\S_{1,1}\times_\vf S^1]$. In this construction, the final step can be implemented by performing the gauging operation on the 3-dimensional theory living on the $\vf$-duality wall of the 4-dimensional theory\cite{Gang:2013sqa,Terashima:2011qi}. The gauging data $\vec{k}\in(\mathbb{Z}^*)^n$ can be read off from the decomposition of the monodromy $\vf=ST^{k_1}ST^{k_2}\cdots ST^{k_n}$ where $S$ and $T$ are ${\rm SL}(2,\mathbb{Z})$ generators defined as
\begin{align}
    S\equiv \begin{pmatrix}
        0&1\\
        -1&0
    \end{pmatrix}\;,\quad T\equiv\begin{pmatrix}
        1&0\\
        1&1
    \end{pmatrix}\;.
\end{align}
The resulting theory is the 3-dimensional $\CN=4$ $\BS(\vec{k})$ theory, whose IR fixed point is identified with $T[\S_{1,1}\times_\vf S^1]$, up to a decoupled unitary TQFT. We provide more details in section~\ref{subsec: generalized S-fold SCFT}.
\paragraph{IR equivalence} The IR equivalence among these theories can be summarized as
\begin{align}
    T[\S_{1,1}\times_\vf S^1]\sim T_{\rm DGG}[\vf]\sim \BS(\vec{k};\vf=ST^{k_1}ST^{k_2}\cdots ST^{k_n})
\end{align}
where $\sim$ indicates that the IR fixed points of the two theories are identical up to a decoupled unitary TQFT. We want to emphasize that the topology of $\S_{1,1}\times_\vf S^1$ depends only on the conjugacy class of $\vf$, and so does the theory $T[\S_{1,1}\times_\vf S^1]$. For the $\BS(\vec{k})$ theory, field theory analysis suggests the existence of a decoupled unitary TQFT ${\rm TFT}[\vec{k}]$ that depends only on $\vec{k}$. If two distinct choices $\vec{k}$ and $\vec{k}'$ correspond to monodromies in the same conjugacy class $\{\vf\}$, we expect the stronger equivalence relation
\begin{align}
    \frac{\BS(\vec{k})}{{\rm TFT}[\vec{k}]}\equiv \CS\{\vf\}\cong \frac{\BS(\vec{k}')}{{\rm TFT}[\vec{k}']}
\end{align}
where $\cong$ denotes equivalence between the IR fixed points of the two theories up to \textit{minimal topological operations} that preserve the absolute values of partition functions on arbitrary closed 3-manifolds, as introduced in (2.27) of \cite{Gang:2024tlp}. We introduce the notation $\CS\{\vf\}$ to highlight this expectation.

\subsection{Abelian Chern-Simons-matter theory} \label{subsec: TDGG}
\begin{figure}[h]
\centering
\includegraphics[width=.60\textwidth]{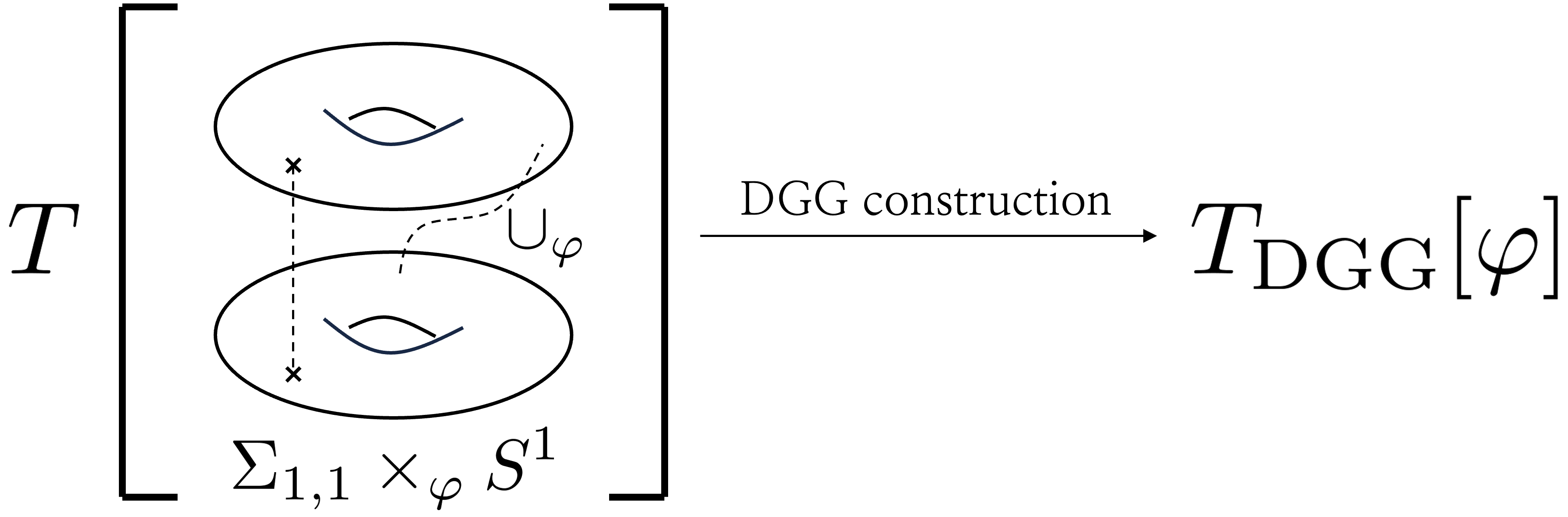}
\captionof{figure}{Abelian Chern-Simons-matter description($T_{\rm DGG}$) of the $T[\S_{1,1}\times_\vf S^1]$ theory. The ideal triangulation $\CT$ of $\S_{1,1}\times_\vf S^1$ is assumed to be provided by the Farey tessellation determined by the decomposition $\vf=\vf(L,R)$. See \eqref{eq: Def of L and R} below for the definitions of $L$ and $R$. We denote the resulting theory by $T_{\rm DGG}[\vf=\varphi(L,R)]$.}
\label{fig: TDGG construction}
\end{figure}
\noindent A generic monodromy $\vf\in{\rm SL}(2,\mathbb{Z})$ can be expressed as a product of $L$ and $R$ matrices
\begin{align} \label{eq: Def of L and R}
    L\equiv\begin{pmatrix}
        1&1\\0&1
    \end{pmatrix}\;,\quad R\equiv \begin{pmatrix}
        1&0\\
        1&1
    \end{pmatrix}\;.
\end{align}
The Farey tessellation\cite{Gu_ritaud_2006} then provides an ideal triangulation of $\S_{1,1}\times_\vf S^1$, which serves as input to the construction. In this paper, we focus on the cases when $\vf=LR^{n}$ and $\vf=L^2R^n$. Following the dictionary developed in~\cite{Dimofte:2011ju,Dimofte:2011py}, we obtain $T_{\rm DGG}[\vf=L^{m}R^n]$ for $m=1,2$. A detailed procedure for the $m=1$ case, including an example of the Farey tessellation, can be found in Appendix A of \cite{Gang:2023ggt}. Here, we only summarize the resulting $T_{\rm DGG}[\vf=L^mR^n]$. We use the \textit{tetrahedron theory} $T_\D$, defined as\cite{Dimofte:2011ju,Dimofte:2011py}
\begin{align}
    T_\D \equiv \text{Chiral }\Phi\text{ with the background Chern-Simons term of level }-\frac{1}{2} \;,
\end{align}
to describe the matter content of the theory. In every case, we are able to introduce a superpotential that breaks the topological symmetries associated with the multiple ${\rm U}(1)$s down to a single flavor ${\rm U}(1)$. This ${\rm U}(1)$ is identified with the ${\rm U}(1)_A$ appearing in the $\CN=2$ decomposition of the $\CN=4$ supersymmetry, which emerges through SUSY enhancement along the RG flow\cite{Gang:2023ggt}.
\paragraph{When $\vf=LR^n$} As shown in \cite{Gang:2023ggt}, the DGG construction gives
\begin{align}
\begin{split}
    &T_{\rm DGG}[\vf=LR^n]
    \\&=\left[{\rm U}(1)_K^{\otimes n+1}\text{ coupled to } n+1\text{ } T_{\D_{i=1,2,\cdots, n+1}} \text{ with }\CW_{\rm sup}=\sum_{m=1}^{n}\CV_m\right]\;.
\end{split}
\end{align}
The gauge charges are diagonal. The $i$-th chiral is charged only under ${\rm U}(1)^i$, with charge $2$ for $\Phi_1$ and charge $1$ for the other chirals. The Chern-Simons level matrix $K$ is
\begin{align}
\label{eq: L1Rn CS level}
    K=\begin{pmatrix}
        2n & 0 & 2 & 4 & \cdots & 2(n-1)\\
        0 & 2 & 2 & 2 & \cdots & 2\\
        2 & 2 & 4 & 4 & \cdots & 4\\
        4 & 2 & 4 & 6 & \cdots & 6\\
        \vdots & \vdots & \vdots & \vdots & \ddots & \vdots\\
        2(n-1) & 2 & 4 & 6 & \cdots & 2n
    \end{pmatrix}\;.
\end{align}
The superpotential $\CW_{\rm sup}$ is given by a linear combination of $n+1$ gauge invariant 1/2 BPS chiral primary operators, as follows:
\begin{align}
    \begin{split}
        \CV_1&=\begin{cases}
            V_{(-1,-2,2)} & \text{when }n=2\\
            V_{(-1,-1,\mathbf{0}_{n-3},-1,2)} & \text{when }n\geq 3
        \end{cases}\;,\\
        \CV_2 & =V_{(0,2,-1,\mathbf{0}_{n-2})}\f_1\;,\\
        \CV_{m\geq 3} & = V_{(\mathbf{0}_{m-2},-1,2,-1,\mathbf{0}_{n-m})}\;.
    \end{split}
\end{align}
Here, $\mathbf{0}_j\equiv(\overbrace{0,0,\cdots,0}^{j\text{ times}})$ and $V_{\vec{m}}$ denotes the 1/2 BPS bare monopole operator that carries magnetic flux $\vec{m}$. The superpotential breaks the topological $\mathrm{U}(1)_T^{\otimes n+1}$ symmetry down to a single $\mathrm{U}(1)_A$, generated by a linear combination of the $\mathrm{U}(1)_T$ generators
\begin{align}
\label{eq: L1Rn A}
    J^A=\vec{A}\cdot\vec{J}^T\text{ where }\vec{A}=(n,1,2,\cdots,n)\;.
\end{align}

\paragraph{When $\vf=L^2R^n$} The DGG construction gives
\begin{align}
\begin{split}
    &T_{\rm DGG}[\vf=L^2R^n]
    \\&=\left[{\rm U}(1)_K^{\otimes n+2}\text{ coupled to }n+2\text{  }T_{\D_{i=1,2,\cdots, n+2}}\text{ with }\CW_{\rm sup}=\sum_{m=1}^{n+1} \CV_m\right]\;.
\end{split}
\end{align}
The gauge charges are diagonal. The $i$-th chiral is charged only under ${\rm U}(1)^i$, with charge $2$ for $\Phi_2$ and charge $1$ for the other chirals. The Chern-Simons level matrix $K$ is
\begin{align}
\label{eq: L2Rn CS level}
    K=\begin{pmatrix}
        1 & -2 & -1 & -1 & \cdots & -1\\
        -2 & 2n+4 & 2 & 4 & \cdots & 2n\\
        -1 & 2 & 2 & 2 & \cdots & 2\\
        -1 & 4 & 2 & 4 & \cdots & 4\\
        \vdots & \vdots & \vdots & \vdots & \ddots & \vdots\\
        -1 & 2n & 2 & 4 & \cdots & 2n
    \end{pmatrix}\;.
\end{align}
The superpotential $\CW_{\rm sup}$ is given by a linear combination of two($n+2$) gauge invariant 1/2 BPS chiral primary operators when $n=1$($n\ge2$), as follows:
\begin{align} \label{eq: L2Rn A charge}
    \begin{split}
        \CV_{1} & =V_{(0,1,\mathbf{0}_{n-1},-1)}\f_1\;,\\
        \CV_{2} & =V_{(2,\mathbf{0}_{n+1})} \f_2^2 \f_3^2\cdots \f_{n+2}^2\;,\\
        \CV_{3} & =\begin{cases}
            V_{(-2,-1,-2,2)} & \text{when }n=2\\
            V_{(-2,-1,-1,\mathbf{0}_{n-3},-1,2)} & \text{when }n\geq 3
        \end{cases}\;,\\
        \CV_{4} & =V_{(0,0,2,-1,\mathbf{0}_{n-2})} \f_2\;,\\
        \CV_{m\ge 5} & =V_{(\mathbf{0}_{m-3},-1,2,-1,\mathbf{0}_{n+2-m})}\;.
    \end{split}
\end{align}
The superpotential breaks the topological ${\rm U}(1)_T^{\otimes n+2}$ symmetry down to a single ${\rm U}(1)_A$, generated by a linear combination of the ${\rm U}(1)_T$ generators
\begin{align}
\label{eq: L2Rn A}
    J^A=\vec{A}\cdot \vec{J}^T\text{ where }\vec{A}=(0,n,1,2,\cdots,n)\;.
\end{align}

\subsection{Generalized S-fold SCFT}
\label{subsec: generalized S-fold SCFT}
\begin{figure}[h]
\centering
\includegraphics[width=.60\textwidth]{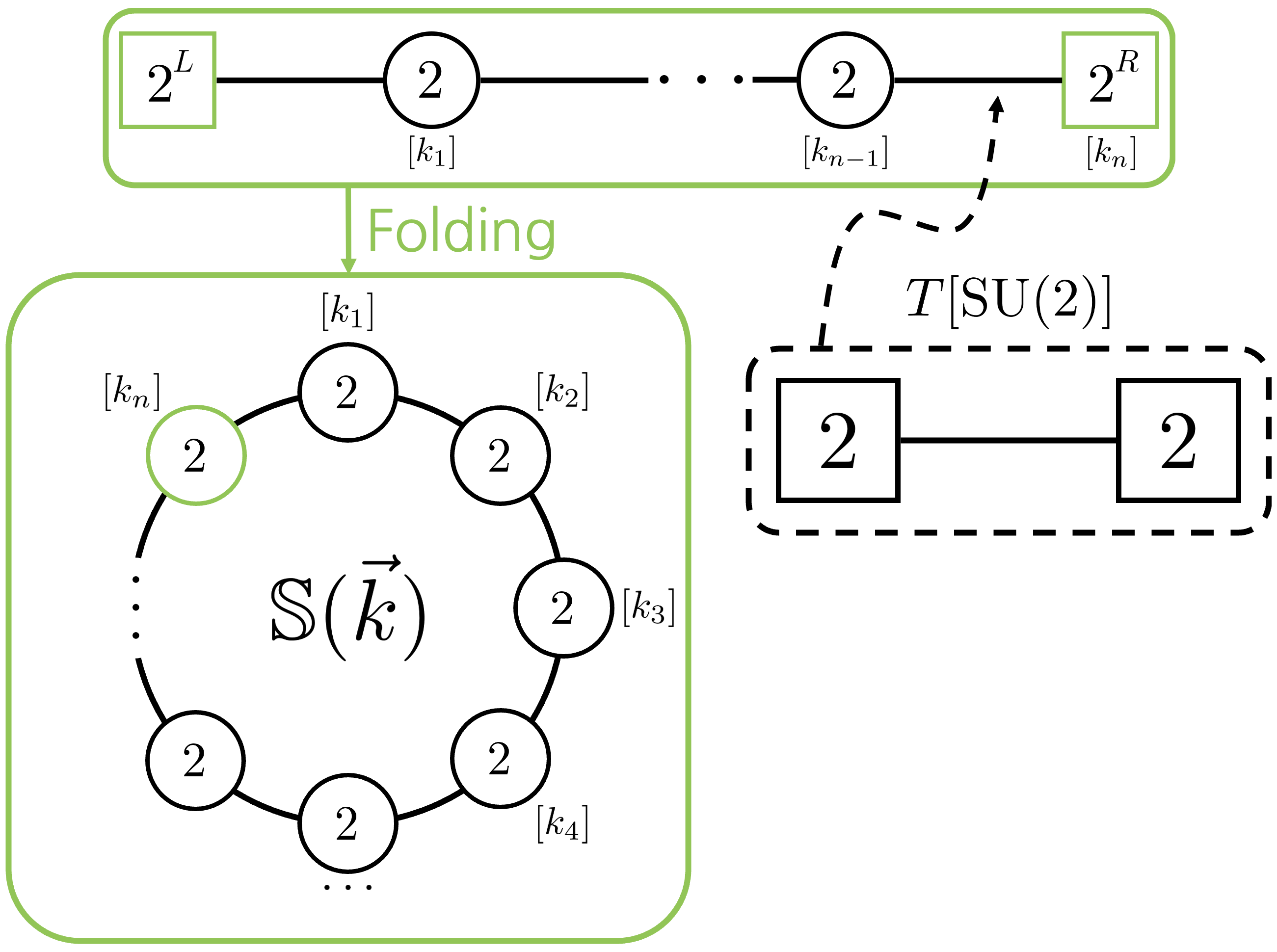}
\captionof{figure}{Construction of the $\BS(\vec{k})$ theory.}
\label{fig: S(k) theory construction}
\end{figure}
\noindent The ${\rm SL}(2,\mathbb{Z})$ duality($S$-duality) of 4-dimensional $\CN=4$ supersymmetric Yang-Mills theory with gauge group $G={\rm SU}(2)$ allows us to consider a class of 3-dimensional $\CN=4$ rank-0 theories labeled by ${\rm SL}(2,\mathbb{Z})$ elements\cite{Gaiotto:2008ak,Gang:2022kpe,Gang:2023ggt}. A generic ${\rm SL}(2,\mathbb{Z})$ elements can be expressed as a product of the two generators $S$ and $T$, which satisfy
\begin{align}
    S^2=(ST)^3=C\;,\quad C^2=1\;.
\end{align}
The actions of $S$ and $T$ on the Yang-Mills complex coupling $\t$ are given by
\begin{align}
    S:\t\longmapsto-\frac{1}{\t}\;,\quad T:\t\longmapsto\t+1\;.
\end{align}
Then for ${\rm SL}(2,\mathbb{Z})$ element $\vf=ST^{k_1}ST^{k_2}\cdots ST^{k_n}$ where $\vec{k}\in(\mathbb{Z}^*)^n$, we can naturally associate the 3-dimensional supersymmetric field theory by considering the theory living on the duality wall, which preserves half of the supersymmetry\footnote{In terms of 3-dimensional supersymmetry, this corresponds to $\CN=4$.}\cite{Gaiotto:2008ak}. The UV description of the theory can be summarized as a linear quiver diagram shown in figure~\ref{fig: S(k) theory construction}. The theory has an ${\rm SU}(2)^{\otimes 2}$ flavor symmetry and gauged ${\rm SU}(2)^{\otimes n-1}$, inherited from the $T[{\rm SU}(2)]$ theories\footnote{The pure $S\in{\rm SL}(2,\mathbb{Z})$ transformation corresponds to the $T[{\rm SU}(2)]$ theory\cite{Gaiotto:2008ak}.} represented by line segments connecting adjacent nodes. The two flavor symmetries, ${\rm SU}(2)^L$ and ${\rm SU}(2)^R$, are represented by the two boxed nodes at the ends of the quiver. Each circular node represents a diagonally gauged ${\rm SU}(2)$, while $[k_i]$ denotes the $\CN=3$ Chern-Simons term at level $k_i$ for the corresponding ${\rm SU}(2)$. Then the $\BS(\vec{k})$ theory we are interested in can be obtained by gauging the diagonal subgroup of the ${\rm SU}(2)^L\otimes {\rm SU}(2)^R$
\begin{align}
\begin{split}
    &\text{Generalized S-fold SCFT }\BS(\vec{k})\equiv\frac{{T[{\rm SU}(2)]^{\otimes n}}}{{\rm SU}(2)_{k_1}^1\otimes{\rm SU}(2)^2_{k_2}\otimes\cdots\otimes{\rm SU}(2)^n_{k_n}}\;,\quad \vec{k}\in(\mathbb{Z}^*)^n\;,
    \\&{\rm SU}(2)^i:\text{the diagonal subgroup of the two SU}(2)\text{s associated with adjacent }T[{\rm SU}(2)]\text{s}\;.
\end{split}
\end{align}
The manifest supersymmetry in the UV description is $\CN=3$, rather than $\CN=4$, due to the presence of the Chern-Simons term. However, the nilpotency of the moment map operator induced in the IR\footnote{Thus, the fundamental identity\cite{Gaiotto:2008sa,Gaiotto:2008sd} is satisfied, and we can consider theories on the ${\rm SL}(2,\mathbb{Z})$ orbit\cite{Gaiotto:2008ak}.} makes the Chern-Simons term also invariant under the $\CN=4$ supersymmetry\cite{Gaiotto:2008ak,Gang:2022kpe}. Moreover, gauging each ${\rm SU}(2)$ at a nonzero Chern-Simons level is expected to completely lift the associated Coulomb- and Higgs-branch operators\cite{Gang:2022kpe,Garozzo:2018kra}. This expectation has been checked through computations of the topologically twisted index\cite{Benini:2015noa} for other theories constructed in the same manner\cite{Gang:2024tlp,Gang:2022kpe,Gang:2023ggt,Jeong:2025xid}.

The $\BS(\vec{k})$ theory has a $(\mathbb{Z}_2)^{\otimes n}$ 1-form symmetry originating from the gauged ${\rm SU}(2)^{\otimes n}$. The associated 't Hooft anomaly can be described by the 4-dimensional anomaly action\cite{Hsin:2018vcg}
\begin{align} \label{eq: 4D anomaly theory}
    \pi\int_{\CM_4}\sum_{i=1}^n\left(\frac{k_i}{2}\CP\left(\CB^i\right)+\CB^i\cup \CB^{i+1}\right)\;,\quad \CB^{n+1}\equiv \CB^1\;.
\end{align}
The 2-form gauge fields $\CB^i \in H^2(\CM_4,\mathbb{Z}_2)$ are introduced for each $\mathbb{Z}_2$ of ${\rm SU}(2)^i$ and $\CP$ is the Pontryagin square operation\cite{Hsin:2018vcg}. This suggests the existence of a decoupled TQFT that realizes the same anomaly in the IR\cite{Hsin:2018vcg}. We denote it as ${\rm TFT}[\vec{k}]$ and define the $\CS\{\vf\}$ theory as the residual sector obtained by quotienting out ${\rm TFT}[\vec{k}]$
\begin{align} \label{eq: Theory Decoupling}
    \BS(\vec{k})\equiv \CS\{\vf\}\otimes{\rm TFT}[\vec{k}]\;.
\end{align}
A candidate for ${\rm TFT}[\vec{k}]$ is the ${\rm U}(1)^{\otimes n}_{\CK}$ Chern-Simons theory with a level matrix that satisfies\cite{Hsin:2018vcg,Gang:2024tlp}
\begin{align} \label{eq: TFT CS matrix}
    \frac{\CK_{ij}}{2}\equiv \bigr({\rm K}_+[\vec{k}]\bigr)_{ij}\;({\rm mod}\;2)\;.
\end{align}
See \eqref{eq: Global linear ansatz coefficient matrix} for the definition of ${\rm K}^+[\vec{k}]$. From \eqref{eq: TFT CS matrix}, we can estimate the dimension of the decoupled TQFT using the fact that the dimension of the ${\rm U}(1)^{\otimes n}_\CK$ theory is bounded above by $\det\CK$ when $\CK$ is non-degenerate\cite{Gang:2024tlp}. Following the same logic in appendix B of~\cite{Gang:2024tlp}, we expect
\begin{align} \label{eq: dim expect}
    \dim {\rm TFT}[\vec{k}]\begin{cases}
        =2^n&\text{when }p_\pm\text{ is odd}\;,
        \\ <2^n&\text{when }p_\pm\text{ is even}\;.
    \end{cases}
\end{align}

\section{Characters of non-unitary Haagerup-like RCFTs}
\label{sec: abelian CSM theory}
\subsection{Construction of a TQFT via topological twisting}
The R-symmetry group of the 3-dimensional $\CN=4$ theory is ${\rm SO}(4)_R\cong{\rm SU}(2)_C\times {\rm SU}(2)_H$. The subscripts $C$ and $H$ stand for \textit{Coulomb} and \textit{Higgs}, respectively. From this, we can identify two independent $\mathfrak{su}(2)\cong\mathfrak{so}(3)$-valued R-connections that can be identified with the spin connection acting on the supercharges. Using either $\mathfrak{su}(2)_C$ or $\mathfrak{su}(2)_H$, we can construct a supercharge $Q_T$ that is independent of the background metric. This procedure is widely known as the \textit{topological twist}\cite{Rozansky:1996bq}. As a consequence, the observables in the $Q_T$-invariant subsector become independent of the metric, thereby defining a topological quantum field theory\cite{Rozansky:1996bq}.

For a concrete construction, it is useful to decompose the ${\rm SO}(4)_R$ R-symmetry in terms of the $\CN=2$ algebra. The two Cartan generators of ${\rm SU}(2)_C$ and ${\rm SU}(2)_H$ are identified with the ${\rm U}(1)_R$ and ${\rm U}(1)_A$ generators through the relations
\begin{align}
    J^R=J^C+J^H\;,\quad J^A=J^C-J^H\;.
\end{align}
Here, ${\rm U}(1)_R$ corresponds to the superconformal R-symmetry in the IR\cite{Gang:2021hrd,Jafferis:2010un}. Since ${\rm U}(1)_A$ acts as a flavor symmetry, it can be mixed with the ${\rm U}(1)_R$ R-symmetry from the $\CN=2$ point of view. We can therefore introduce a mixing parameter $\n$ that labels the choice of R-symmetry
\begin{align} \label{eq: definition of mixing parameter}
    J^R_\n\equiv J^R+\n J^A=J^C+J^H+\n(J^C-J^H)\;.
\end{align}
The topological twist can be implemented by choosing $\n=\pm 1$, depending on which connection is used. We consider the $\n=-1$ case, which is commonly referred to as the topological A-twist\cite{Gang:2021hrd,Gang:2024tlp,Closset:2026xjj}. For $\CN=4$ rank-0 theories, the absence of local operators in the $Q_T$-invariant subsector is guaranteed, and the theories behave like ordinary TQFTs\cite{Witten:1988hf,Atiyah:1989vu,Reshetikhin:1991tc,Dijkgraaf:1989pz}, although they are generally non-unitary\cite{Gang:2021hrd,Closset:2026xjj,Gang:2022kpe,Gang:2024tlp,Gang:2023ggt}. We perform the topological A-twist on the $T_{\rm DGG}[\vf=L^mR^n]$ introduced in section~\ref{subsec: TDGG} to construct the (non-unitary) TQFT $T_{\rm DGG}[\vf=L^mR^n]|_A$.

\subsection{Simplie lines and RCFT characters}
Since the theory is defined as the sector protected by supersymmetry, various BPS partition functions\cite{Kim:2009wb,Hama:2010av,Hama:2011ea,Benini:2015noa,Pestun:2016zxk,Dimofte:2017tpi,Gang:2009qdj} are useful to extract its data. Certain BPS Wilson lines or their linear combinations in the UV theory flow to simple lines in the (non-unitary) IR TQFT\cite{Gang:2024loa}. Whether a UV line operator flows to a simple line in the TQFT can be determined by computing the RG-invariant superconformal index\cite{Kim:2009wb} with an insertion of the line operator\cite{Witten:1988hf}. On the $S^2\times_q S^1$ background, where the superconformal index is computed, such BPS line operators can be supported on two distinguished circles, located at antipodal points of $S^2$ and wrapping $S^1$ with opposite orientations. We label them by $\pm$. For the ${\rm U}(1)^{\otimes r}_K$ Chern-Simons-matter theories with $r$ chirals that we introduced in section~\ref{subsec: TDGG}, the superconformal index with the insertion of BPS Wilson lines $\CL_a$ on the $S^1_+$ and $\CL_b$ on the $S^1_-$ is computed as\cite{Kapustin:2009kz,Kim:2009wb,Gang:2009qdj,Closset:2019hyt}
\begin{align} \label{eq: SCI for U(1) dual theory}
\begin{split}
    &\left\langle \CL_a^+\CL_b^-\right\rangle_{\rm SCI}(q;\eta,\n)
    \\&=\sum_{\vec{m}\in\BZ^r}\oint\prod_{i=1}^r\frac{d u_i}{2 \p i u_i} \prod_{i,j=1}^r u_i^{K_{ij} m_j} \prod_{i=1}^r\CI_{\D}(Q_{i} m_i,u_i^{Q_{i}}) \h^{\vec{A}\cdot\vec{m}} (-q^{1/2})^{(\vec{\m}_0+\n\vec{A})\cdot\vec{m}}W^{+}_aW^-_b
    \\ &\text{where }W^{\pm}_{a=(\vec{Q}',Q_R)}\equiv (-q^{\pm 1/2})^{Q_R}\prod_{i=1}^r\bigr(q^{m_i/2}u_i^{\pm1}\bigr)^{Q_i'}\;.
\end{split}
\end{align}
$Q_{i}$ is the ${\rm U}(1)^i$ gauge charge of the $i$-th chiral. $\CI_\D(m,u)$ is the tetrahedron index\cite{Dimofte:2011py}. $\vec{A}$ specifies the linear combination of topological symmetry generators corresponding to $\mathrm{U}(1)_A$ as $J^A=\vec{A}\cdot\vec{J}^T$, where $J^A$ is the generator of $\mathrm{U}(1)_A$ and $J_i^T$ is the generator of the $i$-th $\mathrm{U}(1)$ topological symmetry. The fugacity for ${\rm U}(1)_A$ is introduced as $\eta$. $\vec{\m}_0=-\vec{A}$ labels the superconformal R-symmetry and an alternative R-symmetry can be chosen by tuning the mixing parameter $\n$, as defined as~\eqref{eq: definition of mixing parameter}. We can consider various BPS Wilson lines in the UV theory, each labeled by its gauge charge $\vec{Q}'$ and the R-charge $Q_R$. Then, line operators $\{\CO_a\}$ flow to distinct simple lines of the TQFT if\cite{Gang:2024loa,Witten:1988hf}
\begin{align} \label{eq: simple criteria}
    \left\langle\CO_a^+\CO_b^-\right\rangle_{\rm SCI}(q;\eta=1,\n=-1)=\d_{ab}\;.
\end{align}
Here, the line operator $\CO_a$ need not be a single BPS Wilson line; it may instead be a linear combination of BPS Wilson lines.

Given the set of line operators $\{\CO_a\}$ that flow to simple lines in the TQFT at the A-twist point, we can study the \textit{bulk-boundary correspondence}, which states that simple lines in the bulk TQFT correspond to chiral primaries in the boundary RCFT\cite{Witten:1988hf}. A series of works \cite{Zagier:2007knq,Nahm:1992sx,Nahm:1994vas,Nahm:2004ch,Berkovich:1994es} expresses a large class of RCFT characters in terms of the \textit{Nahm sum formula}
\begin{align}
    \chi_{(A,B,C)}(q)\equiv\sum_{\vec{m}\in\mathbb{N}^r}\frac{q^{\frac{1}{2}\vec{m}^T \cdot A\cdot \vec{m}+B^T\cdot \vec{m}+C}}{(q)_{m_1}\cdots (q)_{m_r}}\;,\quad A\succcurlyeq0\;.
\end{align}
Motivated by these works, the simple line-chiral primary correspondence has recently been investigated through half-index computations with simple line insertions\cite{Gang:2024loa,Gang:2025ykf}. The twisted $D^2\times_q S^1$ partition function with the half-BPS boundary condition defines the half-index of the 3-dimensional $\CN=2$ gauge theory\cite{Dimofte:2017tpi}. Under the following choice of boundary conditions
\begin{align}
\begin{split}
    \text{Vectors : }&A_0\pm A_1\bigr|_\pd=0\;,\quad D\bigr|_{\pd}=0\;,\quad \l_-\bigr|_{\pd}=0\;,
    \\ \text{Chirals : }&\phi\bigr|_\pd=c\neq 0\;,\quad \ps_+\bigr|_\pd=0\;,
\end{split}
\end{align}
the half-index of the ${\rm U}(1)^{\otimes r}_K$ Chern-Simons-matter theory with $r$ chirals, with the insertion of a BPS Wilson line $\CL_{a=(\vec{Q}',Q_R)}$ is computed as\cite{Dimofte:2017tpi,Yoshida:2014ssa}
\begin{align}\label{eq: half index def}
    \begin{split}
        & \left\langle\CL_{a=(\vec{Q}',Q_R)}\right\rangle_{\mathrm{half}}(q;\h,\n)\\
        & = \sum_{\vec{m}\in\BN_{0}^r} \frac{q^{\frac{1}{2}\vec{m}\cdot K\cdot\vec{m}}}{(q)_{Q_1 m_1} (q)_{Q_2 m_2} \cdots (q)_{Q_r m_r}} \h^{-\vec{A}\cdot\vec{m}} (-q^{1/2})^{-(\vec{\m}_0+\n \vec{A})\cdot\vec{m}} (-q^{1/2})^{Q_R} q^{-\vec{Q}'\cdot\vec{m}}
    \end{split}
\end{align}
where $(q)_n=\prod_{i=1}^n (1-q^i)$ is the Pochhammer symbol. It is proposed that the RCFT character of the chiral primary corresponding to a simple line $\CO_a$ can be computed as\cite{Dimofte:2017tpi, Gang:2023rei,Gang:2024loa,Gang:2025ykf}
\begin{align}\label{eq: half index character}
    \chi_a(q)=q^{h_a-\frac{c}{24}}\langle \CO_a\rangle_{\rm half}(q;\eta=1,\n=-1)\;.
\end{align}
Here, $c$ is the central charge of boundary RCFT, and $h_a$ is the conformal weight of the chiral primary. Equations \eqref{eq: half index def} and \eqref{eq: half index character} allow us to express the characters in closed form as
\begin{align} \label{eq: computed characters form}
    \chi_a(q)=q^{h_a-\frac{c}{24}}\sum_{\vec{m}\in\mathbb{N}^r_0}\frac{q^{\frac{1}{2}\vec{m}^T\cdot K\cdot\vec{m}+\vec{A}\cdot\vec{m}}}{(q)_{Q_1m_1}(q)_{Q_2m_2}\cdots (q)_{Q_rm_r}}\tilde{\chi}_a(q)\;,
\end{align}
where $\tilde{\c}_a(q)$ is determined by the corresponding simple line. In the following subsections, we propose the set of line operators $\{\CO_a\}$ which flow to the simple lines in $T_{\rm DGG}[\vf=LR^{n\geq 2}]|_A$ and $T_{\rm DGG}[\vf=L^2R^{n\geq 1}]|_A$. The field theory data are provided in section~\ref{subsec: TDGG}. In particular, the axial charge vector $\vec{A}$ can be found in \eqref{eq: L1Rn A} and \eqref{eq: L2Rn A} for $\vf=LR^{n\ge 1}$ and $\vf=L^2R^{n\ge 1}$, respectively. We use the following vectors to specify the charge vectors of the BPS Wilson lines:
\begin{align}
    u^{(j)}_i\equiv i+\min (i,j-1)\;,\quad v^{(j)}_i\equiv \min (i,n-j+1)\;,\quad \vec{a}\equiv(a,a,\cdots,a)\;.
\end{align}
We also present the conformal data and the distinguished parts $\tilde{\chi}_a(q)$ appearing in the closed form expressions \eqref{eq: computed characters form} for the characters of the corresponding boundary RCFTs. The characters must satisfy the following modularity condition:
\begin{align} \label{eq: modularity}
    \chi_a(q=e^{2\pi i(-1/\t)})=\sum_bS_{ab}\chi_b(q=e^{2\pi i\t})\;.
\end{align}
where $S_{ab}$ is the modular $S$ matrix. We have numerically verified this relation for various values of $q$ in each RCFT.

\subsubsection{$\vf=LR^{n\ge 2}$}
We propose the following set of $(2n+6)$ line operators as a complete set of simple lines $\{\CO_a\}$ satisfying criteria \eqref{eq: simple criteria}:
\begin{align}\label{eq: L1Rn simple lines}
        \begin{split}
            \CO_0 & =\CL_{\vec{0},0}\;,\\
        \CO_1 & =\CL_{(n,\vec{u}^{(0)}),0}-\CL_{(n-2,\vec{u}^{(0)}),0}\;,\\
        \CO_2 & =\CL_{\vec{A},0}\;,\\
        \CO_3 & =\CL_{\vec{A}+(n,\vec{u}^{(0)}),0}-\CL_{\vec{A}+(n-2,\vec{u}^{(0)}),0}\;,\\
        \CO_{3+\a} & =\left.\CL_{(n+\a,\vec{u}^{(\a+2)}-\vec{1}),0}-\CL_{(n+\a-2,\vec{u}^{(\a+1)}-\vec{1}),0}+\CL_{(n-\a,\vec{v}^{(\a)}-\vec{1}),0}\right|_{1\leq \a\leq n-1}\;,\\
        \CO_{n+3} & =\CL_{(n-1,\vec{u}^{(1)}),0}\;,\\
        \CO_{n+4} & =\CL_{(n,\vec{u}^{(2)}),0}-\CL_{(n,\vec{u}^{(1)}),0}\;,\\
        \CO_{n+5} & =\CL_{(n+1,\vec{u}^{(3)}),0}-\CL_{(n+1,\vec{u}^{(1)}),0}-\CL_{(n-1,\vec{v}^{(2)}),0}\;,\\
        \CO_{n+5+\b} & =\left.\CL_{(n+\b-1,\vec{u}^{(\b+1)}+\vec{1}),0}-\CL_{(n+\b-1,\vec{u}^{(\b+1)}),0}-\CL_{(n+\b-3,\vec{u}^{(\b)}),-2}\right|_{1\leq\b\leq n}\;.
    \end{split}
\end{align}
From \eqref{eq: L1Rn simple lines}, the conformal weights of the chiral primaries are determined as
\begin{align}
        \vec{h}=\biggr(0,-\frac{n}{4},-\frac{n}{4},0,\frac{\a^2}{4n}-\frac{n}{4}\biggr|_{1\le \a\le n-1},\frac{\b^2}{4(n+4)}-\frac{n}{4}\biggr|_{1\le \b\le n+3}\biggr)\;,
    \end{align}
 with the central charge $c=-6 n+1$. Here is the list of $\tilde{\chi}_a(q)$:
\begin{align}
        \begin{split}
    \tilde{\chi}_0&=1\;,
    \\\tilde{\chi}_1&=q^{-(n,\vec{u}^{(0)})\dm}-q^{-(n-2,\vec{u}^{(0)})\dm}\;,
    \\\tilde{\chi}_2&=q^{-\vec{A}\dm}\;,
    \\\tilde{\chi}_3&=q^{-\vec{A}\dm}\paren{q^{-(n,\vec{u}^{(0)})\dm}-q^{-(n-2,\vec{u}^{(0)})\dm}}\;,
    \\\tilde{\chi}_{3+\a}&=\left.q^{-(n+\a,\vec{u}^{(\a+2)}-\vec{1})\dm}-q^{-(n+\a-2,\vec{u}^{(\a+1)}-\vec{1})\dm}+q^{-(n-\a,\vec{v}^{(\a)}-\vec{1})\dm}\right|_{1\leq \a\leq n-1} \;,
    \\\tilde{\chi}_{n+3}&=q^{-(n-1,\vec{u}^{(1)})\dm}\;,
    \\\tilde{\chi}_{n+4}&=q^{-(n,\vec{u}^{(2)})\dm}-q^{-(n,\vec{u}^{(1)})\dm}\;,
    \\\tilde{\chi}_{n+5}&= q^{-(n+1,\vec{u}^{(3)})\dm}-q^{-(n+1,\vec{u}^{(1)})\dm}-q^{-(n-1,\vec{v}^{(2)})\dm}\;,
    \\\tilde{\chi}_{n+5+\b}&=\left.q^{-(n+\b-1,\vec{u}^{(\b+1)}+\vec{1})\dm}-q^{-(n+\b-1,\vec{u}^{(\b+1)})\dm}-q^{-1} q^{-(n+\b-3,\vec{u}^{(\b)})\dm}\right|_{1\leq \b\leq n}\;.
\end{split}
    \end{align}
Their $q$-series expansions obtained from the closed form expression \eqref{eq: computed characters form} agree with the RCFT characters given in section 3.3 of \cite{Gang:2023ggt}. We also check that the modularity condition \eqref{eq: modularity} is satisfied by the following modular $S$ matrix:
     \begin{align} \label{eq: mod S LRn}
        \begin{split}\begin{pmatrix}
        \begin{matrix}
            (-1)^n a_0 & a_0\\
            a_0 & a_0
        \end{matrix} & \vline & \begin{matrix}
            a_3 & (-1)^n a_3\\
            a_3 & a_3
        \end{matrix} & \vline & \begin{matrix}
            -a_1 & a_1 & \cdots & (-1)^{n-1} a_1\\
            a_1 & a_1 & \cdots & a_1
        \end{matrix} & \vline & \begin{matrix}
            a_2 & -a_2 & \cdots & (-1)^{n+2} a_2\\
            -a_2 & -a_2 & \cdots & -a_2
        \end{matrix}\\
        \hline
        \begin{matrix}
            a_3 & a_3\\
            (-1)^n a_3 & a_3
        \end{matrix} & \vline & \begin{matrix}
            a_0 & a_0\\
            a_0 & (-1)^n a_0
        \end{matrix} & \vline & \begin{matrix}
            a_1 & a_1 & \cdots & a_1\\
            -a_1 & a_1 & \cdots & (-1)^{n-1} a_1
        \end{matrix} & \vline & \begin{matrix}
            a_2 & a_2 & \cdots & a_2\\
            -a_2 & a_2 & \cdots & (-1)^{n+3} a_2
        \end{matrix}\\
        \hline
        \begin{matrix}
            -a_1 & a_1\\
            a_1 & a_1\\
            \vdots & \vdots\\
            (-1)^{n-1} a_1 & a_1
        \end{matrix} & \vline & \begin{matrix}
            a_1 & -a_1\\
            a_1 & a_1\\
            \vdots & \vdots\\
            a_1 & (-1)^{n-1} a_1
        \end{matrix} & \vline & 2 a_1 \left.\cos\paren{\frac{\p \a \a'}{n}}\right|_{1\leq \a,\a'\leq n-1} & \vline & \mathbf{0}\\
        \hline
        \begin{matrix}
            a_2 & -a_2\\
            -a_2 & -a_2\\
            \vdots & \vdots\\
            (-1)^{n+2} a_2 & -a_2
        \end{matrix} & \vline & \begin{matrix}
            a_2 & -a_2\\
            a_2 & a_2\\
            \vdots & \vdots\\
            a_2 & (-1)^{n+3} a_2
        \end{matrix} & \vline & \mathbf{0} & \vline & 2 a_2 \left.\cos\paren{\frac{\p \b \b'}{n+4}}\right|_{1\leq \b,\b'\leq n+3}
    \end{pmatrix}
        \end{split}
    \end{align}
    where
    \begin{align}
    (a_0,a_1,a_2,a_3)\equiv\biggr(\frac{1}{2}\bigr(\frac{1}{\sqrt{2n}}+\frac{1}{\sqrt{2(n+4)}}\bigr),\frac{1}{\sqrt{2n}},\frac{1}{\sqrt{2(n+4)}},\frac{1}{2}\bigr(\frac{1}{\sqrt{2n}}-\frac{1}{\sqrt{2(n+4)}}\bigr)\biggr)\;.
\end{align}

\subsubsection{$\vf=L^2R^{n\ge 1}$}
We propose the following set of $(2n+4)$ line operators as a complete set of simple lines $\{\CO_a\}$ satisfying criteria \eqref{eq: simple criteria}:
\begin{align}\label{eq: L2Rn simple lines}
        \begin{split}
            \CO_0 & =\CL_{\vec{0},0}\;,\\
        \CO_1 & =\CL_{(-1,n+2,\vec{u}^{(1)}),0}-\CL_{(-1,n,\vec{u}^{(1)}),0}\;,\\
        \CO_2 & =\CL_{\vec{A},0}\;,\\
        \CO_3 & =\CL_{\vec{A}+(-1,n+2,\vec{u}^{(1)}),0}-\CL_{\vec{A}+(-1,n,\vec{u}^{(1)}),0}\;,\\
        \CO_{3+\a} & =\left.\CL_{(-1,n+\a+2,\vec{u}^{(\a+1)}),0}-\CL_{(-1,n+\a,\vec{u}^{(\a+1)}),0}-\CL_{(0,n-\a,\vec{v}^{(\a+1)}),-1}\right|_{1\leq \a\leq n-1}\;,\\
        \CO_{n+2+\b} & =\left.\CL_{(-1,n+\b,\vec{u}^{(\b)}),0}-\CL_{(-1,n+\b-2,\vec{u}^{(\b)}),0}+\CL_{(0,n-\b,\vec{v}^{(\b)}),0}\right|_{1\leq\b\leq n+1}\;.
        \end{split}
\end{align}
From \eqref{eq: L2Rn simple lines}, the conformal weights of the chiral primaries are determined as
\begin{align}
        \vec{h}=\biggr(0,-\frac{n}{4}+\frac{1}{2},-\frac{n}{4},\frac{1}{2},\frac{\a^2}{4n}-\frac{n}{4}+\frac{1}{2}\biggr|_{1\le \a\le n-1},\frac{\b^2}{4(n+2)}-\frac{n}{4}\biggr|_{1\le \b\le n+1}\biggr)\;.
    \end{align}
with the central charge $c=-6 n+2$.
Here is the list of $\tilde{\chi}_a(q)$:
\begin{align}
        \begin{split}
            \tilde{\chi}_0&=1\;,
        \\\tilde{\chi}_1&=q^{-(-1,n+2,\vec{u}^{(1)})\dm}-q^{-(-1,n,\vec{u}^{(1)})\dm}\;,
        \\\tilde{\chi}_2&=q^{-\vec{A}\dm}\;,
        \\\tilde{\chi}_3&=q^{-\vec{A}\dm} \paren{q^{-(-1,n+2,\vec{u}^{(1)})\dm}-q^{-(-1,n,\vec{u}^{(1)})\dm}}\;,
        \\\tilde{\chi}_{3+\a}&=\left.q^{-(-1,n+\a+2,\vec{u}^{(\a+1)})\dm}-q^{-(-1,n+\a,\vec{u}^{(\a+1)})\dm}+q^{-1/2} q^{-(0,n-\a,\vec{v}^{(\a+1)})\dm}\right|_{1\leq\a\leq n-1}\;,
        \\\tilde{\chi}_{n+2+\b}&=\left.q^{-(-1,n+\b,\vec{u}^{(\b)})\dm}-q^{-(-1,n+\b-2,\vec{u}^{(\b)})\dm}+q^{-(0,n-\b,\vec{v}^{(\b)})\dm}\right|_{1\leq\b\leq n+1}\;.
        \end{split}
\end{align}
After factoring out $q^{h_a-\frac{c}{24}}$, the remaining $q$-series contain half-integer powers of $q$, indicating that the corresponding RCFTs are fermionic\cite{Gang:2023ggt,Duan:2022kxr,Cho:2022kzf}. We also check that the modularity condition \eqref{eq: modularity} is satisfied by the following modular $S$ matrix:
\begin{align} \label{eq: mod S L2Rn}
        \begin{split}
    \begin{pmatrix}
        \begin{matrix}
            (-1)^n a_0 & a_0\\
            a_0 & a_0
        \end{matrix} & \vline & \begin{matrix}
            a_3 & (-1)^n a_3\\
            a_3 & a_3
        \end{matrix} & \vline & \begin{matrix}
            -a_1 & a_1 & \cdots & (-1)^{n-1} a_1\\
            a_1 & a_1 & \cdots & a_1
        \end{matrix} & \vline & \begin{matrix}
            a_2 & -a_2 & \cdots & (-1)^{n} a_2\\
            -a_2 & -a_2 & \cdots & -a_2
        \end{matrix}\\
        \hline
        \begin{matrix}
            a_3 & a_3\\
            (-1)^n a_3 & a_3
        \end{matrix} & \vline & \begin{matrix}
            a_0 & a_0\\
            a_0 & (-1)^n a_0
        \end{matrix} & \vline & \begin{matrix}
            a_1 & a_1 & \cdots & a_1\\
            -a_1 & a_1 & \cdots & (-1)^{n-1} a_1
        \end{matrix} & \vline & \begin{matrix}
            a_2 & a_2 & \cdots & a_2\\
            -a_2 & a_2 & \cdots & (-1)^{n+1} a_2
        \end{matrix}\\
        \hline
        \begin{matrix}
            -a_1 & a_1\\
            a_1 & a_1\\
            \vdots & \vdots\\
            (-1)^{n-1} a_1 & a_1
        \end{matrix} & \vline & \begin{matrix}
            a_1 & -a_1\\
            a_1 & a_1\\
            \vdots & \vdots\\
            a_1 & (-1)^{n-1} a_1
        \end{matrix} & \vline & 2 a_1 \left.\cos\paren{\frac{\p \a \a'}{n}}\right|_{1\leq \a,\a'\leq n-1} & \vline & \mathbf{0}\\
        \hline
        \begin{matrix}
            a_2 & -a_2\\
            -a_2 & -a_2\\
            \vdots & \vdots\\
            (-1)^{n} a_2 & -a_2
        \end{matrix} & \vline & \begin{matrix}
            a_2 & -a_2\\
            a_2 & a_2\\
            \vdots & \vdots\\
            a_2 & (-1)^{n+1} a_2
        \end{matrix} & \vline & \mathbf{0} & \vline & 2 a_2 \left.\cos\paren{\frac{\p \b \b'}{n+2}}\right|_{1\leq \b,\b'\leq n+1}
    \end{pmatrix}
        \end{split}
    \end{align}
    where
    \begin{align}
    (a_0,a_1,a_2,a_3)\equiv\biggr(\frac{1}{2}\bigr(\frac{1}{\sqrt{2n}}+\frac{1}{\sqrt{2n+4}}\bigr),\frac{1}{\sqrt{2n}},\frac{1}{\sqrt{2n+4}},\frac{1}{2}\bigr(\frac{1}{\sqrt{2n}}-\frac{1}{\sqrt{2n+4}}\bigr)\biggr)\;.
\end{align}

\section{More modular matrices of non-unitary Haagerup-like TQFTs}
\label{sec: generalized S-fold}
In the previous section, we proposed the central charges, conformal weights, modular $S$ matrices and characters of the Haagerup-like RCFTs only for the theories labeled by certain monodromies $\vf$. For general monodromies $\vf$, we cannot determine the complete set of characters because the full set of simple lines is not known. Nevertheless, the Bethe vacua analysis allows us to extract partial information about the modular $S$ and $T$ matrices. In this section, we present a Bethe vacua analysis of the generalized S-fold SCFTs introduced in section~\ref{subsec: generalized S-fold SCFT}. We then propose the modular $S$ and $T$ matrices of the non-unitary TQFT for a broader class of monodromies $\vf$.

\paragraph{Parameters} In this section, parameters that depend on the trace of the monodromy $\vf$ appear repeatedly in the modular matrices. We denote them by
\begin{align}
    p_{\pm}\equiv \mathrm{tr} \vf\pm 2\;.
\end{align}

\subsection{Bethe vacua analysis} \label{subsec: S(k) BG analysis}
Following the dictionary developed in \cite{Gang:2021hrd}, we perform the Bethe vacua analysis\cite{Gang:2021hrd,Closset:2019hyt,Nekrasov:2014xaa} to extract partial modular data of the $\BS(\vec{k})|_A$ theory. The overall procedure goes as follows. The squashed 3-sphere partition function is defined as the Euclidean path integral on the squashed 3-sphere background labeled by the squashing parameter $b^2$:
\begin{align} \label{eq: squashed sphere}
    S^3_b\equiv\left\{x\in\mathbb{R}^4\;\left| \;b^2\bigr((x^1)^2+(x^2)^2\bigr)+\frac{1}{b^2}\bigr((x^3)^2+(x^4)^2\bigr)=1\right.\right\}\;.
\end{align}
Through supersymmetric localization, the path integral over the dynamical fields is converted to an ordinary integral over the real mass parameters of the vector multiplets, labeling their BPS configurations\cite{Pestun:2016zxk,Hama:2010av,Hama:2011ea}. The resulting partition function depends on two parameters\footnote{The real mass parameter associated with ${\rm U}(1)_A$ is set to zero to preserve $\CN=4$ supersymmetry\cite{Terashima:2011qi}.}, $\n$ and $b$, where $\n$ is the mixing parameter between ${\rm U}(1)_A$ and ${\rm U}(1)_R$.

Since we are interested in the theory at $\n=-1$, where metric independence emerges, we can choose a convenient $b^2$ to evaluate the integral. It has proven useful to consider the $b^2\ra0$ limit. In this limit, the integral admits a saddle point approximation due to the smallness of $\hb\equiv 2\pi ib^2$. By introducing collective notation $\vec{U}$ for ($2\pi b$-rescaled) real mass parameters of the dynamical ${\rm U}(1)(Z_{i=1,\cdots,n})$ and ${\rm SU}(2)(X_{i=1,\cdots, n})$ vector multiplets, the integral becomes
\begin{align} \label{eq: Integral in squashing limit}
\begin{split}
    Z^{S^3_b}_{\BS(\vec{k})|_A}\xrightarrow{b^2\ra0}&\int \frac{d^{2n}U}{(2\pi \hb)^n}\exp\left.\left[\frac{1}{\hb}\CW^{\BS(\vec{k})}_0+\CW_1^{\BS(\vec{k})}+\CO(\hb)\right]\right|_{\n=-1}
\end{split}
\end{align}
and the modified saddle point equation (Bethe equation) $\exp\bigr(\pd_{\vec{U}}\CW_0\bigr)=1$ together with physical equivalence relations, defines the set of Bethe vacua\cite{Closset:2019hyt,Nekrasov:2014xaa,Gang:2021hrd}. For each ${\rm SU}(2)^i$, the associated ${\mathbb{Z}}_2$ Weyl subgroup acts on the solution $V$ as $X_i\longmapsto-X_i$\footnote{Consequently, the ${\rm U}(1)$ real mass parameters are transformed as well.} and one can identify the $2^n$ Weyl-related solutions by collecting solutions that have different signs for the ${\rm SU}(2)$ real mass parameters but share the same physical $\HF$ datum, which we will define shortly. By taking the quotient of the solutions by the Weyl subgroup of the gauge group\footnote{It can be done by choosing one solution from among the $2^n$ Weyl-related solutions\cite{Gang:2021hrd,Jeong:2025xid,Gang:2024tlp}.} and the equivalence relation $V_i\sim V_i+2\pi i$, we find the Bethe vacua
\begin{align}
    \BV\equiv\biggr\{\;V\;\biggr|\;e^{\pd_{\vec{U}}\CW_0^{\BS(\vec{k})}}\bigr|_{\n=-1,\vec{U}=V}=1\;\biggr\}\biggr/\bigr((\textbf{Weyl}),\;V_i\sim V_i+2\pi i\bigr)\;.
\end{align}
Then, for each Bethe vacuum $V$, we can extract the $\HF$ datum from the $1$-loop($\hb^0$) data and the classical phase($\hb^{-1}$) of the integral \eqref{eq: Integral in squashing limit} around $\vec{U}=V$. The pair of Handle-gluing $\CH(V)$ and Fibering $\CF(V)$\cite{Closset:2019hyt,Nekrasov:2014xaa}
\begin{align} \label{eq: HF datum}
\begin{split}
    \CH(V)&\equiv\frac{e^{i\d}}{2^{2n}}\det_{i,j}\left[-\pd_{U_i}\pd_{U_j}\CW_0^{\BS(\vec{k})}\right]e^{-2\CW_1^{\BS(\vec{k})}}\biggr|_{\n=-1,\vec{U}=V}\;,
    \\\CF(V)&\equiv \exp\left[-\frac{1-\vec{U}\cdot\pd_{\vec{U}}}{2\pi i}\CW_0^{\BS(\vec{k})}\right]_{\n=-1,\vec{U}=V}
\end{split}
\end{align}
defines a single $\HF$ datum. Here, the phase factor $e^{i\d}$ in the Handle-gluing can be determined from the condition $\CH>0$. By collecting them for all Bethe vacua $\BV$, we find the $\HF$ data that allow us to evaluate various BPS partition functions. We can check that the sum of the inverse Handle-gluing is $1$. This supports the claim that the $\BS(\vec{k})$ theory is rank-0 through its local operator counting interpretation on the radially quantized Hilbert space with the R-charge fugacity\cite{Benini:2015noa,Closset:2019hyt}
\begin{align}
    \sum_{\HF^{\BS(\vec{k})}}\CH^{-1}={\rm Tr}_{\CH(S^2)}(-1)^R=1\;.
\end{align}
The 3-sphere partition function of the theory can also be evaluated from the $\HF$ data\cite{Closset:2019hyt}. As a consequence, the dictionary developed in~\cite{Gang:2021hrd} states that the absolute values of the entries in the first row of the modular $S$ matrix and the modular $T$ matrix can be found from the $\HF$ data and the $|S_{00}|$ can be determined since it should be equal to the 3-sphere partition function of the TQFT
\begin{align} \label{eq: S3 ptf from HF sum}
\begin{split}
    &\HF^{\BS(\vec{k})}\equiv \bigr\{(\CH,\CF)\bigr\}=\bigr\{(|S_{0a}|^{-2},\x T_{aa})\bigr\}\;,\quad \x:\text{Phase}
    \\&\longrightarrow\left|\sum_{\HF^{\BS(\vec{k})}}\CH^{-1}\CF\right|=\left|Z_{\BS(\vec{k})|_A}^{S^3_b}\right|=|S_{00}|\;.
\end{split}
\end{align}
Note that \eqref{eq: S3 ptf from HF sum} implies we can find $\{V^0\}\subset\BV$ satisfying $\left|\bigr(\CH(V^0)\bigr)\right|^{-\frac{1}{2}}=\left|Z_{\BS(\vec{k})|_A}^{S^3_b}\right|$. There exists a unique element in $\{V^0\}$ mapped to the trivial simple line $\CO_0\equiv\mathbf{1}$ through the \textit{Bethe vacuum-Simple line map}\cite{Gang:2021hrd} which allows us to determine the full modular $S$ matrix when we know the map for all Bethe vacua\cite{Gang:2021hrd,Jeong:2025xid}
\begin{align} \label{eq: S matrix from simple lines}
    V^a\longleftrightarrow\CO_a\text{ then }S_{ab}=\z\cdot\CO_b[V^a]\;.
\end{align}
The constant $\z$ can be determined from physical conditions that the modular matrix $S$ should satisfy\cite{XGWen,Gannon:2003de,Verlinde:1988sn}. We specify these conditions and use them to determine the matrix in the next section~\ref{subsec: HI from S(k)}.

Now we provide the analysis specific to the A-twisted $\BS(\vec{k})$ theory we are interested in. We cannot strictly set the value of $\n$ to $-1$ a priori since the contribution of the adjoint chiral in each $T[{\rm SU}(2)]$ segment diverges at $\n=-1$. To address this problem, it has been proven to be useful to introduce a small parameter $\e\equiv 1+\n$ and define the observables in the limit $\e\ra0$\cite{Jeong:2025xid}. The divergence of observables can be avoided by an appropriate limiting behavior of the Bethe vacuum. The \textit{global linear ansatz} considered in~\cite{Jeong:2025xid}
\begin{align} \label{eq: Global linear ansatz}
    Z_i=(-1)^{G_i}X_i+f_i\e+\CO(\e^2)\;,\quad \left.X_i\right|_{\e=0}\notin \pi i\mathbb{Z}\;,\quad G_i\in\mathbb{Z}_2
\end{align}
still works. In addition to~\eqref{eq: Global linear ansatz}, we can also consider the \textit{local linear ansatz} specific to the site index $i=1,2,\cdots,n$. As a representative example, we study the $n=1$ and $n=2$ cases in detail. Here, we display Bethe vacua and associated $\HF$ data only. A more detailed procedure, including the full solutions before taking the Weyl quotient, can be found in appendix~\ref{sec: Detail BG analysis}. The results for the $n=1$ cases are in agreement with previous work \cite{Gang:2022kpe,Gang:2023ggt}, providing a consistency check for the ansatz approach.

\paragraph{When $n=1$} Note that $p_\pm=k_1\pm 2$. We find Bethe vacua labeled by $\bar{\a}\in\{1,2,\cdots,|p_-|-1\}$ and $\bar{\b}\in\{1,2,\cdots,|p_+|-1\}$ from the global linear ansatz \eqref{eq: Global linear ansatz}
\begin{align}
    \label{eq: n=1 Global linear BV}
    \begin{split}
        {\rm BV}^{\BS(k_1)}_{\rm GL-} & \equiv\biggr\{\;(Z_1=-X_1,X_1)\;\biggr|\;X_1=\frac{\p i}{p_-}\bar{\a}\;\biggr\}\;,
        \\{\rm BV}^{\BS(k_1)}_{\rm GL+} & \equiv\biggr\{\;(Z_1=X_1,X_1)\;\biggr|\;X_1=\frac{\p i}{p_+}\bar{\b}\;\biggr\}\;.
    \end{split}
\end{align}
The associated $\HF$ data are evaluated as
\begin{align}
    \begin{split}
        \HF^{\BS(k_1)}_{\rm GL-} & \equiv\biggr\{\paren{2|p_-|,e^{\frac{5\p i}{12}+\frac{\p i}{2 p_-}\bar{\a}^2}}\;\biggr|\;\bar{\a}\in\{1,2,\cdots,|p_-|-1\}\biggr\}\;,
        \\\HF^{\BS(k_1)}_{\rm GL+} & \equiv\biggr\{\paren{2|p_+|,e^{\frac{5\p i}{12}+\frac{\p i}{2 p_+}\bar{\b}^2}}\;\biggr|\;\bar{\b}\in\{1,2,\cdots,|p_+|-1\}\biggr\}\;.
    \end{split}
\end{align}
When $p_\pm$ is \textit{odd}, the $\HF$ data factorized to
\begin{align}
\begin{split}
    \HF^{\BS(k_1)}_{\rm GL-}&=\biggr\{\paren{|p_-|,e^{\frac{5\pi i}{12}+\frac{2\pi i}{p_-}\a^2}}\;\biggr|\;\a\in\bigr\{1,2,\cdots,\frac{|p_-|-1}{2}\bigr\}\biggr\}\otimes\HF^{{\rm TFT}[k_1]}\;,
    \\\HF^{\BS(k_1)}_{\rm GL+}&=\biggr\{\paren{|p_+|,e^{\frac{5\pi i}{12}+\frac{2\pi i}{p_+}\b^2}}\;\biggr|\;\b\in\bigr\{1,2,\cdots,\frac{|p_+|-1}{2}\bigr\}\biggr\}\otimes \HF^{{\rm TFT}[k_1]}
\end{split}
\end{align}
with the decoupled $\HF^{{\rm TFT}[k_1]}\equiv\bigr\{(2^1,1),(2^1,e^{\frac{\pi i}{2}(k_1+2)})\bigr\}$.

\paragraph{When $n=2$} We assume both $k_1$ and $k_2$ are \textit{odd integers} and $|k_1k_2-4|>2$, so $p_\pm$ are odd and $p_+p_->0$. We find Bethe vacua labeled by $\a_i\in \bigr\{1,2,\cdots,\frac{|k_i|-1}{2}\bigr\}$, $\b\in\bigr\{1,2,\cdots,\frac{|p_-|-1}{2}\bigr\}$, and $s\in\{\pm 1\}$, $\vec{H}\in\mathbb{Z}_2^2$ from the global linear ansatz \eqref{eq: Global linear ansatz}
\begin{align} \label{eq: n=2 Global linear BV}
\begin{split}
    \BV_{\rm GL-}&\equiv\biggr\{\;(Z_i=(-1)^iX_i,X_i)\;\biggr|\;\vec{X}=2\pi i\left(\frac{s\a_1}{k_1},\frac{\a_2}{k_2}\right)+\pi i\vec{H}\;\biggr\}\;,
    \\\BV_{\rm GL+}&\equiv\biggr\{\;(Z_i=X_i,X_i)\;\biggr|\;\vec{X}=\frac{2\pi i}{p_-}(k_2,-2)\b+\pi i\vec{H}\;\biggr\}\;.
\end{split}
\end{align}
The associated $\HF$ data are evaluated as
\begin{align}
\label{eq: n=2 GL HF}
    \begin{split}
        \HF^{\BS(\vec{k})}_{\rm GL-}&=\left\{\left.\left(|p_+|,e^{\frac{5\pi i}{6}+\frac{2\pi i}{k_1}\a_1^2+\frac{2\pi i}{k_2}\a_2^2}\right)\right|\a_i\in\bigr\{1,2,\cdots,\frac{|k_i|-1}{2}\bigr\}\right\}^{\otimes 2}\otimes\HF^{{\rm TFT}[\vec{k}]}\;,
        \\\HF^{\BS(\vec{k})}_{\rm GL+}&=\left\{\left.\left(|p_-|,e^{\frac{5\pi i}{6}+\frac{2k_2\pi i}{p_-}\b^2}\right)\right|\b\in\bigr\{1,2,\cdots,\frac{|p_-|-1}{2}\bigr\}\right\}\otimes\HF^{{\rm TFT}[\vec{k}]}\;
    \end{split}
\end{align}
with the decoupled $\HF^{{\rm TFT}[\vec{k}]}\equiv\left\{\left.\left(2^2,e^{\frac{\pi i}{2}\vec{k}\cdot\vec{H}}\right)\right|\vec{H}\in\mathbb{Z}_2^2\right\}$.

In addition to \eqref{eq: n=2 Global linear BV}, we find Bethe vacua labeled by $\a_i\in\bigr\{1,2,\cdots,\frac{|k_i|-1}{2}\bigr\}$ and $\vec{H}\in\mathbb{Z}_2^2$ from the local linear ansatz
\begin{align}
\begin{split}
    \BV_{(A,D)}&\equiv\biggr\{\;(Z_i,X_i)\;\biggr|\;\vec{X}=\pi i\left(\frac{2\a_1}{k_1}+H_1,H_2\right),\;Z_1=\log\left[\frac{1+i\sinh X_1}{\cosh X_1}\right],\;Z_2=X_2\;\biggr\}\;,
    \\ \BV_{(D,A)}&\equiv\biggr\{\;(Z_i,X_i)\;\biggr|\;\vec{X}=\pi i\left(H_1,\frac{2\a_2}{k_2}+H_2\right),\;Z_1=X_1,\;Z_2=\log\left[\frac{1+i\sinh X_2}{\cosh X_2}\right]\;\biggr\}\;.
\end{split}
\end{align}
Here, the subscripts $(A,D)$ and $(D,A)$ label the local ansatze used. See appendix~\ref{subsec: Local ansatz appendix} for their definition. Interestingly, the resulting $\vec{X}$ can be viewed as the $\a_i=0$ extension of the Bethe vacua \eqref{eq: n=2 Global linear BV} from the global linear ansatz, while one of the $Z_i$ becomes a \textit{real number}. The associated $\HF$ data are evaluated as
\begin{align}
\label{eq: n=2 LL HF}
\begin{split}
    \HF^{\BS(\vec{k})}_{(A,D)}&=\left\{\left.\left(|p_+|,e^{\frac{5\pi i}{6}+\frac{2\pi i}{k_1}\a_1^2}\right)\right|\a_1\in\bigr\{1,2,\cdots,\frac{|k_1|-1}{2}\bigr\}\right\}\otimes\HF^{{\rm TFT}[\vec{k}]}\;,
    \\ \HF^{\BS(\vec{k})}_{(D,A)}&=\left\{\left.\left(|p_+|,e^{\frac{5\pi i}{6}+\frac{2\pi i}{k_2}\a_2^2}\right)\right|\a_2\in\bigr\{1,2,\cdots,\frac{|k_2|-1}{2}\bigr\}\right\}\otimes\HF^{{\rm TFT}[\vec{k}]}
\end{split}
\end{align}
with the decoupled $\HF^{{\rm TFT}[\vec{k}]}$ being the same as that obtained from the global linear ansatz.

\paragraph{Global square root ansatz} The A-twisted $\BS(\vec{k})$ theory admits a \textit{global square root ansatz} of the form
\begin{align}
     Z_i=\pi iH_i+g_i^Z\e^{1/2}\;,\quad X_i=\pi iH_i+g_i^X\e^{1/2}\;,\quad H_i\in\mathbb{Z}_2
\end{align}
in the $\e\ra0$ limit. For this class of Bethe vacua, both the solutions and the associated degrees of freedom can be easily tracked for arbitrary $n$. After taking the Weyl quotient, we find Bethe vacua labeled by $s\in\{\pm 1\}$ and $\vec{H}\in\mathbb{Z}_2^n$
\begin{align} \label{eq: GS BV}
    \BV_{\rm GS}\equiv\bigr\{\pi i(\vec{H},\vec{H})\bigr\}^{\otimes 2}\;.
\end{align}
Note that $s=\pm 1$ labels the two distinct solutions when $\e\neq 0$. We must regard them as two independent solutions even though they are \textit{degenerate} at $\e=0$. The associated $\HF$ data are evaluated as
\begin{align} \label{eq: HF data of sqrt ansatz} \begin{split}
    &\HF^{\BS(\vec{k})}_{\rm GS}=\left\{\left.\left(2^{-2}p_+p_-\bigr(|p_+|+|p_-|+2s\sqrt{p_+p_-}\bigr),e^{\frac{5n\pi i}{12}}\right)\right|s\in\{\pm 1\}\right\}\otimes\HF^{{\rm TFT}[\vec{k}]}
    \\&\text{with the decoupled }\HF^{{\rm TFT}[\vec{k}]}\equiv \left\{\left.\left(2^n,e^{\frac{\pi i}{2}\vec{H}^T\cdot{\rm K}^+[\vec{k}]\cdot\vec{H}}\right)\right|\vec{H}\in\mathbb{Z}_2^n\right\}\;.
\end{split}\end{align}
We expect that the same decoupled sector appears for generic $n$ when $p_\pm$ is \textit{odd}, regardless of the choice of the ansatz. When $n=2$, we can rewrite $\frac{\pi i}{2}\vec{H}^T\cdot{\rm K}^+[\vec{k}]\cdot\vec{H}\longrightarrow \vec{k}\cdot\vec{H}$ thanks to the $2\pi i$-periodicity and $\HF^{{\rm TFT}[\vec{k}]}$ is identified with the decoupled $\HF$ data that appeared earlier.

By combining $\HF$ data \eqref{eq: HF data of sqrt ansatz} from the global square root ansatz with the other $\HF$ data found for $n=1,2$, we can evaluate the 3-sphere partition function of the $\BS(\vec{k})|_A$ theory from \eqref{eq: S3 ptf from HF sum}. We check that the result agrees with the one computed from the direct integral shown in appendix~\ref{sec: Partition function integral}. The latter computation can be carried out for arbitrary $n$. Since the Handle-gluing corresponding to the $s=-1$ sector of \eqref{eq: HF data of sqrt ansatz} reproduces the 3-sphere partition function
\begin{align}
    |\CH|^{-\frac{1}{2}}=\frac{1}{2^{\frac{n}{2}+1}}\bigr(\frac{1}{\sqrt{|p_+|}}+\frac{1}{\sqrt{|p_-|}}\bigr)=\left|Z^{S^3_b}_{\BS(\vec{k})|_A}\right|\;,
\end{align}
we expect that, when $p_\pm$ is \textit{odd}, the $2^n$ Bethe vacua in the $s=-1$ sector of \eqref{eq: HF data of sqrt ansatz} to map, through the Bethe vacuum-Simple line map, to simple lines of the decoupled ${\rm TFT}[\vec{k}]$ tensored with the trivial simple line of $\CS\{\vf\}|_A$.

\paragraph{Comment on the cases with even $p_\pm$} Regardless of the parity of $p_\pm$, the $\HF$ data from the global square root ansatz have the $2^n$ dimensional decoupled sector
\begin{align}
    \HF[\vec{k}]=\left\{\left(2^n,e^{\frac{\pi i}{2}\vec{H}^T\cdot{\rm K}^+\cdot\vec{H}}\right)\;\biggr|\;\vec{H}\in\mathbb{Z}^n_2\right\}\;.
\end{align}
However, it \textit{cannot} be regarded as the $\HF$ data of a $2^n$ dimensional \textit{bosonic} TQFT when $p_\pm$ is \textit{even}. We can check
\begin{align}
    \left|\sum_{\HF[\vec{k}]}\CH^{-1}\CF\right|=\begin{cases}
        \CH^{-\frac{1}{2}}&(p_\pm\text{ is odd})\;,
        \\\sqrt{2}\CH^{-\frac{1}{2}}\text{ or }0 &(p_\pm\text{ is even})\;.
    \end{cases}
\end{align}
This supports the prediction for the dimension of the decoupled TQFT shown in \eqref{eq: dim expect}. We expect some of the $2^n$ degrees of freedom associated with the $(\mathbb{Z}_2)^{\otimes n}$ 1-form symmetry of gauged ${\rm SU}(2)^{\otimes n}$
\begin{align} \label{eq: Z21fs action}
    (Z_i,X_i)\longmapsto (Z_i+\pi i,X_i+\pi i)
\end{align}
to be \textit{trivialized} when $p_\pm$ is even, because this symmetry is identified with the action of the Weyl subgroup, which provides an equivalence relation between Bethe vacua. We can check it explicitly for the Bethe vacua when $n=1$ shown in \eqref{eq: n=1 global linear full BV}. Since choosing different $W$ is equivalent to adding $\pi i$ for the $\bar{\b}=\frac{|p_+|}{2}(\bar{\a}=\frac{|p_-|}{2})$ element, we \textit{cannot} regard the degrees of freedom in \eqref{eq: Z21fs action} as independent of the Weyl degrees of freedom. Consequently, the $\HF$ data are not factorized when $p_\pm$ is even. In terms of the 't Hooft anomaly, this implies the $\mathbb{Z}_2$ 1-form symmetry is \textit{non-anomalous} and the decoupled TQFT for the anomaly matching is not required. See \cite{Gang:2023ggt,Gang:2022kpe} for a detailed analysis when $n=1$.

\subsection{Modular matrix proposal for $\vf=ST^{k_1\in 2\BZ+1}ST^{k_2\in 2\BZ+1}$} \label{subsec: HI from S(k)}
Through the Bethe vacua analysis in the previous section~\ref{subsec: S(k) BG analysis}, we find the decoupling expected in section~\ref{subsec: generalized S-fold SCFT} in agreement with the structure~\eqref{eq: Theory Decoupling}
\begin{align}
    \underbrace{\bigoplus_{\rm Ansatz}\HF^{\BS(\vec{k})}_{\rm Ansatz}}_{\HF^{\BS(\vec{k})}}\equiv \HF^{\CS\{\vf\}}\otimes \HF^{{\rm TFT}[\vec{k}]}\;.
\end{align}
Each decoupled sector can be interpreted as the HF data of an individual TQFT. For the cases studied in the previous section~\ref{subsec: S(k) BG analysis}, we can check
\begin{align}
    \sum_{\HF}\CH^{-1}=1\;,\quad \exists (\CH_0,\CF_0)\in\HF\text{ such that }\left|\sum_{\HF}\CH^{-1}\CF\right|=|\CH_0|^{-\frac{1}{2}}
\end{align}
and for ${\HF}^{\CS\{\vf\}}$, $(\CH_0,\CF_0)$ is uniquely determined as $s=-1$ element of $\HF^{\BS(\vec{k})}_{\rm GS}/\HF^{\rm TFT[\vec{k}]}$ shown in \eqref{eq: HF data of sqrt ansatz} when $p_\pm$ is \textit{odd}. It is natural to interpret $\HF^{\CS\{\vf\}}$ as $\HF$ data of the (non-unitary) TQFT $\CS\{\vf\}|_A$ defined as the residual sector after modding out the decoupled (unitary) TQFT from the A-twisted $\BS(\vec{k})$ theory. Since the full TQFT $\BS(\vec{k})|_A$ consists of the two decoupled TQFTs, we expect the modular matrices to factorize
\begin{align}
    S^{\BS(\vec{k})}=S^{\CS\{\vf\}}\otimes S^{{\rm TFT}[\vec{k}]}\;,\quad T^{\BS(\vec{k})}=T^{\CS\{\vf\}}\otimes T^{{\rm TFT}[\vec{k}]}
\end{align}
where each sector can be \textit{partially} recovered from the associated $\HF$ data, through the dictionary \eqref{eq: S3 ptf from HF sum}. See \cite{Gang:2021hrd,Jeong:2025xid} for examples. However, conditions imposed on the modular matrices are strong enough that we can \textit{propose} the matrix elements explicitly. In addition to the ${\rm SL}(2,\mathbb{Z})$ defining relations
\begin{align} \label{eq: SL(2,Z) defining relations}
    S^2=(ST)^3=C\;,\quad C^2=1\;,
\end{align}
the modular $S$ matrix should also satisfy the following conditions arising from the physics of the (non-unitary) RCFT expected to be supported on the boundary of the bulk (non-unitary) TQFT
\begin{itemize}
    \item The modular $S$ matrix should contain a \textit{positive row}\cite{Gannon:2003de} where all elements are positive numbers. For a unitary theory, this should be the first row and $S_{00}\le S_{0a}$ should hold\cite{XGWen}.
    \item The \textit{fusion coefficient} $N^a_{bc}$ computed from the modular $S$ matrix should satisfy\cite{XGWen,Verlinde:1988sn}
    \begin{align} \label{eq: fusion coefficient from S matrix}
        N^a_{bc}=\sum_{d}\frac{S_{bd}S_{cd}S_{ad}^*}{S_{0d}}\in\mathbb{N}_0\;.
    \end{align}
\end{itemize}
The proposed modular matrices for the $n=1$ cases can be found in (2.18) of \cite{Gang:2022kpe}. In this paper, we propose modular matrices for a broader class in the $n=2$ case with $p_\pm\in 2\BZ+1$.

\paragraph{When $n=2$} We propose the following modular $S$ matrix:
\begin{align}
    \label{eq: n=2 full S matrix}
    S^{\CS\{\vf\}}=\begin{pmatrix}
        \begin{matrix}
            a_0 & a_3 \\
            a_3 & a_0
        \end{matrix} & \vline & \begin{matrix}
            -\s a_1 &  \cdots & -\s a_1\\
            a_1 &  \cdots & a_1
        \end{matrix} & \vline & \begin{matrix}
            \s a_2 & \cdots & \s a_2\\
            a_2 &  \cdots & a_2
        \end{matrix}\\
        \hline
        \begin{matrix}
            -\s a_1 & a_1\\
            \vdots & \vdots \\
            -\s a_1 & a_1
        \end{matrix} & \vline & \begin{matrix}
            2a_1E^{(1)}&2a_1\mathbf{1}&2a_1D^{(1)}\\
            2a_1\mathbf{1}^T&2a_1E^{(2)}&2a_1D^{(2)}\\
            2a_1D^{(1)T}&2a_1D^{(2)T}&2a_1E^{(3)}
        \end{matrix} & \vline & \mathbf{0}\\
        \hline
        \begin{matrix}
            \s a_2 & a_2\\
            \vdots & \vdots\\
            \s a_2 & a_2
        \end{matrix} & \vline & \mathbf{0} & \vline & \left.2 a_2 \cos\bigr(\frac{4 \p k_2 \b\b'}{p_-}\bigr)\right|_{1\leq \b,\b'\leq \frac{|p_-|-1}{2}}
    \end{pmatrix}\;.
\end{align}
Parameters are defined as $\s\equiv {\rm sign}(p_\pm)$ and
\begin{align}
    (a_0,a_1,a_2,a_3)&\equiv\biggr(\frac{1}{2}\bigr(\frac{1}{\sqrt{|p_+|}}+\frac{1}{\sqrt{|p_-|}}\bigr),\frac{1}{\sqrt{|p_+|}},\frac{1}{\sqrt{|p_-|}},\frac{1}{2}\bigr|\frac{1}{\sqrt{|p_+|}}-\frac{1}{\sqrt{|p_-|}}\bigr|\biggr)
\end{align}
so the second row is a positive row. Entries of submatrices appearing in the middle block are
\begin{align}
    \begin{split}
        \mathbf{1}_{\a _1\a_2}& \equiv1\;,\\
        E^{(1)}_{\a_1,\a_1'}&\equiv\cos\bigr(\frac{4\pi \a_1\a_1'}{k_1}\bigr)\;,\\
        E^{(2)}_{\a_2,\a_2'}& \equiv\cos\bigr(\frac{4\pi \a_2\a_2'}{k_2}\bigr)\;,
        \\D^{(1)}_{\a_1,(s,\a_1',\a_2)}&\equiv\cos\bigr(s\frac{4\pi \a_1\a_1'}{k_1}\bigr)\;\\
        D^{(2)}_{\a_2,(s,\a_1,\a_2')}& \equiv\cos\bigr(\frac{4\pi \a_2\a_2'}{k_2}\bigr)\;,
        \\E^{(3)}_{(s,\a_1,\a_2),(s',\a_1',\a_2')}&\equiv\cos\bigr(ss'\frac{4\pi \a_1\a_1'}{k_1}+\frac{4\pi \a_2\a_2'}{k_2}\bigr)\;
    \end{split}
\end{align}
with $\a_1,\a_1'\in\bigr\{1,2,\cdots,\frac{|k_1|-1}{2}\bigr\}$, $\a_2,\a_2'\in\bigr\{1,2,\cdots,\frac{|k_2|-1}{2}\bigr\}$ and $s,s'\in\{\pm 1\}$. With the modular $T$ matrix:
\begin{align}
\begin{split}
    &T^{\CS\{\vf\}}=e^{-\frac{2\p i c}{24}}{\rm diag}\left[1,1,\vec{T}^{(1)},\vec{T}^{(2)},\vec{T}^{(3)},\left.\exp\bigr(\frac{2\pi ik_2\b^2}{p_-}\bigr)\right|_{1\le \b\le \frac{|p_-|-1}{2}}\right]\text{ where}
    \\&T^{(1)}_{\a_1}\equiv\exp\bigr(\frac{2\pi i\a_1^2}{k_1}\bigr)\;,\quad T^{(2)}_{\a_2}\equiv\exp\bigr(\frac{2\pi i\a_2^2}{k_2}\bigr)\;,\quad T^{(3)}_{(s,\a_1,\a_2)}\equiv\exp\bigr(\frac{2\pi i\a_1^2}{k_1}+\frac{2\pi i\a_2^2}{k_2}\bigr)\;,
\end{split}
\end{align}
we can check that the conditions for the modular matrices are satisfied. The \textit{central charge} of the boundary RCFT\cite{Gannon:2003de} is identified with $c$ in the definition of the modular $T$ matrix, modulo $8$. A nonzero value of $c$ modulo $8$ occurs when $|k_1|\not\equiv |k_2|\;({\rm mod}\;4)$. In that case, there is a unique $k_j$ that satisfies $|k_j|\equiv 3\;({\rm mod}\;4)$. Then $c=2\times{\rm sign}(k_j)\;({\rm mod}\;8)$.

\begin{table}[h]
\centering
\begin{tabular}{|c|c|}
\hline
 Modular matrices of $\CD^\w{\rm Hg}_{2n+1}$& Modular matrices of ${\rm Gal}_p\left.\CS\{\vf=ST^{k_1} ST^{k_2}\}\right|_A$ \\ \hline
 $(\w,n)=(0,1)$& $(p;k_1,k_2)=(2;-3,-3)$ \\
  $(3,4)$& $(31;-3,27)$ \\
 $(2,4)$&$(31;5,17)$  \\
 $(0,2)$&$(2;-5,5)$  \\
$(0,3)$ &$(2;-7,7)$  \\
$(0,4)$ &$(2;-9,9)$  \\
    $\vdots$& $\vdots$ \\
 $(0,n)$&$(2;-2n-1,2n+1)$\\\hline
\end{tabular}
\caption{Comparison between $\CD^{\w}\mathrm{Hg}_{2n+1}$ in \cite{Evans:2010yr} and the modular matrices obtained by Galois conjugation of $\left.\CS\{\vf\}\right|_A$.}
\label{tab: Identification with generalized Haagerup}
\end{table}
For the smallest $N\in\mathbb{N}$ satisfying $T^N\propto \BI$, the \textit{Galois conjugation} of the modular data is defined by an integer $p$ satisfying ${\rm gcd}(p,N)=1$\cite{Harvey:2018rdc}. Under the Galois conjugation, modular matrices are transformed as
\begin{align} \label{eq: Galois conjugation}
    {\rm Gal}_p:\; \bigr(S,T\bigr)\longmapsto \bigr(T^{\Bar p}S^{-1}T^pST^{\Bar p}S^2,T^{\Bar p}\bigr)\;,\quad p\bar{p}\equiv1\;({\rm mod}\;N)\;.
\end{align}
We check that modular matrices of the $\CS\{\vf\}|_A$ theory we propose are identified with the \textit{generalized Haagerup modular data} $\CD^\w {\rm Hg}_{2n+1}$ introduced in section 3.2 of~\cite{Evans:2010yr} through the Galois conjugation~\eqref{eq: Galois conjugation} as shown in table~\ref{tab: Identification with generalized Haagerup}.

\section{Discussion and Future directions}
\paragraph{Behavior of real mass parameter} Except for the Bethe vacua obtained from the local ansatze, the ($2\pi b$-rescaled) real mass parameters of the ${\rm U}(1)$ vector multiplet $Z$ and the ${\rm SU}(2)$ vector multiplet $X$ corresponding to the same $T[{\rm SU}(2)]$ segment are identified up to a sign. The same phenomenon occurs in similarly constructed theories studied in the following works: \cite{Gang:2024tlp,Jeong:2025xid}. It would be interesting to investigate whether this phenomenon can be understood from a UV field theory analysis.

\paragraph{Bethe vacua analysis when ${\rm tr}\vf\in2\mathbb{Z}$} After expanding the Bethe equations \eqref{eq: BE without ansatz} to third order in $\e$ using the $(A,D)$-type (or $(D,A)$-type) local ansatz introduced in appendix~\ref{subsec: Local ansatz appendix}, we find that an inconsistency among the equations arises when ${\rm tr}\vf$ is even. This is why we assume that both $k_1$ and $k_2$ are odd integers. It would be interesting to find a limiting behavior(i.e., an ansatz) for the Bethe vacua that allows us to analyze the cases in which ${\rm tr}\vf$ is even. We also expect that understanding the cases in which $\mathrm{tr}\vf$ is even would allow us to enlarge the solvable sector to arbitrary $n\ge 1$. If this procedure can be carried out successfully, our proposed modular $S$ matrix \eqref{eq: mod S L2Rn} for $\vec{k}\in(-2,\mathbb{N})$ would provide a strong consistency check.

\paragraph{Characters of the boundary RCFT for general $\vf$} Finding the complete set of simple lines in the dual abelian description is a highly nontrivial task. To bypass this issue, the so-called Riemann-Hilbert method, developed in a series of works \cite{Bantay:2005vk,Bantay:2007zz,Gannon:2013jua,Cheng:2020srs} was used in previous work \cite{Gang:2023ggt} to construct the complete set of RCFT characters. The method requires not only the modular $S$ and $T$ matrices but also some known characters in order to \textit{uniquely} determine the complete set of RCFT characters. We expect that combining these approaches will be useful for developing a method to determine the full set of characters for arbitrary $\vf$. Enlarging the solvable sector of the $\mathbb{S}(\vec{k})|_A$ theories and identifying \textit{universal} simple lines\footnote{We know at least two of them: the trivial one $\CL_{\vec{0},0}$ and the \textit{axial} simple line $\CL_{\vec{A},0}$ whose gauge charge is identified with the ${\rm U}(1)_A$ charge vector\cite{Gang:2025ykf,Gang:2024loa}.} in their dual abelian descriptions would allow us to determine the complete set of boundary RCFT characters for generic $\vf$.

\paragraph{Simple lines of the $\mathbb{S}(\vec{k})$ theory} For the similarly constructed theories studied in \cite{Gang:2024tlp,Jeong:2025xid}, integer powers of a single principal root of unity $e^{\frac{\pi i}{p}}$ generate the entire set of exponentiated Bethe vacua. Here, $p$ is the determinant of the coefficient matrix appearing in a global linear ansatz such as~\eqref{eq: Global linear ansatz coefficient matrix}. This led the authors to expect that the Wilson lines of ${\rm SU}(2)^i$, in the $(|p|-1)$-th symmetric representation correspond to the simple lines of the decoupled TQFT\cite{Hsin:2018vcg}. This expectation was confirmed, and the modular matrices of the decoupled TQFT could be constructed from the non-abelian description using the dictionary~\eqref{eq: S matrix from simple lines}. In contrast, we find that the exponentiated Bethe vacua of the $\mathbb{S}(\vec{k})$ theory can be written in terms of integer powers of two principal roots of unity, $e^{\frac{\pi i}{p_+}}$ and ${e^{\frac{\pi i}{p_-}}}$. This makes it difficult to determine which Wilson lines correspond to the decoupled TQFT, which is expected to have identical Handle-gluings based on the HF data analysis. Identifying the complete set of simple lines of the $\mathbb{S}(\vec{k})$ theory would allow us to check whether the modular matrices proposed in section~\ref{subsec: HI from S(k)} are correct by constructing the full modular $S$ matrix through the dictionary~\eqref{eq: S matrix from simple lines} and examining its decoupling. We leave this as an interesting direction for future work.

\acknowledgments
We are grateful to Dongmin Gang for his early collaboration and useful discussions. We also thank Heeyeon Kim, Sungjoon Kim and Sungjay Lee for useful discussions. This work was supported in part by the National Research Foundation of Korea (NRF) grant NRF-2022R1C1C1011979. We also acknowledge support from the National Research Foundation of Korea (NRF) grant RS-2024-00405629.

\newpage

\appendix
\section{Details of the Bethe vacua analysis} \label{sec: Detail BG analysis}
Near the A-twist point $\n=-1+\e$, the squashed 3-sphere partition function of the generalized S-fold SCFT $\BS(\vec{k})$, in the $b^2\ra 0$ limit, is given by\cite{Pestun:2016zxk,Hama:2010av,Hama:2011ea,Jeong:2025xid}\footnote{We adopt the charge convention for the $T[{\rm SU}(2)]$ theory introduced in section 2.1 of \cite{Jeong:2025xid}.}
\begin{align}
    Z^{S^3_b}_{\BS(\vec{k})}[m=0,\n=-1+\e]=\int\frac{d^nZd^nX}{(2\pi \hb)^n}\exp\left[\frac{1}{\hbar}\mathcal{W}_0^{\BS(\vec{k})}+\mathcal{W}_1^{\BS(\vec{k})}+\CO(\hbar)\right]
\end{align}
where
\begin{align}
    \begin{split} \label{eq: W0 and W1 of the S(k) theory}
        \CW_0^{\BS(\vec{k})}&=n\Li_2(e^{\e\pi i})+\sum_{i=1}^n\biggr[(k_i+1)X_i^2+Z_i^2+2Z_iX_{i+1}+\sum_{\e_{Z,X}\in\{\pm1\}}\Li_2\bigr(e^{\e_ZZ_i+\e_XX_i-\frac{1}{2}\e\pi i}\bigr)\biggr]\;,
        \\ \CW_1^{\BS(\vec{k})}&=\frac{n(-1+\e)}{2}\Li_1(e^{\e\pi i})+\sum_{i=1}^n\biggr[\log\sinh X_i+\frac{2-\e}{4}\sum_{\e_{Z,X}\in\{\pm 1\}}\Li_1\bigr(e^{\e_ZZ_i+\e_XX_i-\frac{1}{2}\e\pi i}\bigr)\biggr]\;.
    \end{split}
\end{align}
Here, the real mass parameter $m$ associated with ${\rm U}(1)_A$ is set to zero to preserve $\CN=4$ supersymmetry\cite{Terashima:2011qi}. From \eqref{eq: W0 and W1 of the S(k) theory}, we find the Bethe equations $\exp\bigr(\pt_{\vec{Z},\vec{X}}\CW_0\bigr)=1$ as\cite{Closset:2019hyt,Nekrasov:2014xaa,Gang:2021hrd}
\begin{align} \label{eq: BE without ansatz}
    \begin{split}
        \pd_{Z_i}\CW_0^{\BS(\vec{k})}&,\;\pd_{X_i}\CW_0^{\BS(\vec{k})}\in2\pi i\mathbb{Z}\text{ where}
        \\\pd_{Z_i}\CW_0^{\BS(\vec{k})}&=2Z_i+2X_{i+1}+\log\left[\frac{1-e^{-Z_i+X_i-\frac{\pi i}{2}\e}}{1-e^{Z_i-X_i-\frac{\pi i}{2}\e}}\right]+\log\left[\frac{1-e^{-Z_i-X_i-\frac{\pi i}{2}\e}}{1-e^{Z_i+X_i-\frac{\pi i}{2}\e}}\right]\;,
        \\ \pd_{X_i}\CW_0^{\BS(\vec{k})}&=2Z_{i-1}+2(k_i+1)X_i+\log\left[\frac{1-e^{Z_i-X_i-\frac{\pi i}{2}\e}}{1-e^{-Z_i+X_i-\frac{\pi i}{2}\e}}\right]+\log\left[\frac{1-e^{-Z_i-X_i-\frac{\pi i}{2}\e}}{1-e^{Z_i+X_i-\frac{\pi i}{2}\e}}\right]\;.
    \end{split}
\end{align}
The action of the Weyl subgroup and the $\BZ_2$ 1-form symmetry associated with each $\SU(2)^i$ appear as symmetries of equations~\eqref{eq: BE without ansatz}. The following two transformations map one solution of \eqref{eq: BE without ansatz} to another:
\begin{align} \label{eq: Weyl and Z2 1fs}
    \begin{split}
        \text{Weyl}&:(Z_{i-1},X_{i})\longmapsto (-Z_{i-1},-X_i)\;,
        \\ \mathbb{Z}_2\text{ 1-form}&:(Z_i,X_i)\longmapsto (Z_i+\pi i,X_i+\pi i)\;.
    \end{split}    
\end{align}
We use letters $W_i\in\{\pm1\}$ and $H_i\in\mathbb{Z}$ to label Weyl subgroup and $\mathbb{Z}_2$ 1-form symmetry degrees of freedom in solutions, respectively. By fixing $\vec{W}$, we quotient the solutions by the Weyl subgroup and obtain the Bethe vacua. Given the solutions to \eqref{eq: BE without ansatz}, the associated $\HF$ data, defined in \eqref{eq: HF datum}, can be computed from the following second derivatives
\begin{align} \label{eq: Second derivatives of W0}
\begin{split}
    \pd_{Z_i}^2\CW_0^{\BS(\vec{k})}&=2+\sum_{\e_{Z,X}\in\{\pm 1\}}\frac{1}{e^{\e_Z Z_i+\e_X X_i+\frac{\pi i}{2}\e}-1}\;,
    \\\pd_{X_i}^2\CW_0^{\BS(\vec{k})}&=2(k_i+1)+\sum_{\e_{Z,X}\in\{\pm 1\}}\frac{1}{e^{\e_ZZ_i+\e_XX_i+\frac{\pi i}{2}\e}-1}\;,
    \\\pd_{Z_i}\pd_{X_i}\CW^{\BS(\vec{k})}_0&=\sum_{\e_{Z,X}\in\{\pm 1\}}\frac{\e_Z\e_X}{e^{\e_ZZ_i+\e_XX_i+\frac{\pi i}{2}\e}-1}\;,
    \\\quad \pd_{Z_i}\pd_{X_{i+1}}\CW_0^{\BS(\vec{k})} & =2\;,\quad (\text{Otherwise}) =0
\end{split}
\end{align}
and the modified $\CW_0^{\BS(\vec{k})}$
\begin{align} \label{eq: modified W0}
\left(1-\vec{U}\cdot\pd_{\vec{U}}\right)\CW_0^{\BS(\vec{k})}=\frac{5n\pi^2}{6}-\sum_{i=1}^n\left[k_iX_i^2+2Z_iX_{i+1}\right]\;.
\end{align}
The action of the Weyl subgroup should leave the $\HF$ datum invariant. It appears as the $\vec{W}$-independence of the $\HF$ datum. To solve the Bethe equations \eqref{eq: BE without ansatz}, we introduce three ansatze--the \textit{global linear ansatz}, the \textit{local linear ansatz}, and the \textit{global square root ansatz}--which we discuss in turn below.
\subsection{Global linear ansatz}
We assume that, near the A-twist point $\n=-1+\e$, the solutions take the form
\begin{align}
    Z_i=(-1)^{G_i}X_i+f_i\e+\CO(\e^2)\;,\quad \left.X_i\right|_{\e=0}\notin\pi i\mathbb{Z}\;,\quad G_i\in\mathbb{Z}_2
\end{align}
for $i=1,2,\cdots, n$. We treat the cases with $n\leq 2$ separately. In each case, $\vec{Z}$ at order $\e^0$ is completely determined by $\vec{G}$ and $\vec{X}$.
\subsubsection{Bethe vacua for $n=1$}
For $n=1$, the Bethe equations \eqref{eq: BE without ansatz} reduce to the single equation
\begin{align}
    \bigr(k_1+2(-1)^{G_1}\bigr)X_1\in\pi i\mathbb{Z}\;.
\end{align}
Assuming $|k_1|\geq 3$, there are $2 (|p_+|+|p_-|-2)$ independent solutions labeled by $\bar{\a}\in\{1,2,\cdots,|p_-|-1\}$, $\bar{\b}\in\{1,2,\cdots,|p_+|-1\}$ and $W\in \{\pm 1\}$:
\begin{align} \label{eq: n=1 global linear full BV}
    \begin{split}
        X_1=W \frac{\p i \bar{\a}}{p_-}&\longrightarrow \CH=2|p_-|\;,\quad \CF=\exp\paren{\frac{5\pi i}{12}+\frac{\p i}{2p_-} \bar{\a}^2}\;,\\
        X_1=W \frac{\p i \bar{\b}}{p_+}&\longrightarrow \CH=2|p_+|\;,\quad \CF=\exp\paren{\frac{5\pi i}{12}+\frac{\p i}{2 p_+}\bar{\b}^2}\;.
    \end{split}
\end{align}
$W$ implements the action of the Weyl subgroup. Since the $\HF$ data are independent of the choice of $W$, we may fix $W=1$, thereby quotienting the solutions by the Weyl subgroup. This yields \eqref{eq: n=1 Global linear BV}.

\subsubsection{Bethe vacua for $n=2$ with $(k_1, k_2)\in (2\BZ+1)^2$}
For $n=2$, the Bethe equations \eqref{eq: BE without ansatz} reduce to the matrix equation
\begin{align}
    \begin{pmatrix}
        k_1&(-1)^{G_1}+(-1)^{G_2}\\
        (-1)^{G_1}+(-1)^{G_2}&k_2
    \end{pmatrix}\cdot\vec{X}\in\pi i\mathbb{Z}^2\;.
\end{align}
For the two cases $(-1)^{G_1}=-(-1)^{G_2}=\widetilde{W}$ where $\widetilde{W}=\pm 1$, we find $(|k_1|-1)\times (|k_2|-1)\times 2^2$ independent solutions labeled by $\a_i\in \{1,2,\cdots,\frac{|k_i|-1}{2}\}$, $s\in \{\pm 1\}$, $W\in \{\pm 1\}$ and $\vec{H}\in \BZ_2^2$:
\begin{align}\label{eq: n=2 global linear Km}
\begin{split}
    \vec{X}&=2\pi i W\left(\frac{s\a_1}{k_1},\frac{\a_2}{k_2}\right)+\pi i\vec{H}\;,
    \\\CH&=4|k_1k_2|\;,\quad\CF=\exp\left(\frac{5\pi i}{6}+\frac{2\p i}{k_1} \a_1^2+\frac{2\p i}{k_2}\a_2^2+\frac{\p i}{2}\vec{k}\cdot\vec{H}\right)\;.
\end{split}
\end{align}
For the other two cases $(-1)^{G_1}=(-1)^{G_2}=\widetilde{W}$ where $\widetilde{W}=\pm 1$, we find $(|p_-|-1)\times 2^2$ independent solutions labeled by $\b\in \{1,2,\cdots,\frac{|p_-|-1}{2}\}$, $W\in \{\pm 1\}$ and $\vec{H}\in \BZ_2^2$:
\begin{align} \label{eq: n=2 global linear Kp}
\begin{split}
    \vec{X}&=\frac{2\pi iW}{p_-}\left(k_2,-2\widetilde{W}\right)\b+\pi i\vec{H}\;,
    \\ \CH&=4|k_1k_2-4|\;,\quad\CF=\exp\left(\frac{5\pi i}{6}+\frac{2\p i k_2}{p_-} \b^2+\frac{\p i}{2}\vec{k}\cdot\vec{H}\right)\;.
\end{split}
\end{align}
$(W,\widetilde{W})$ and $\vec{H}$ implement the actions of the Weyl subgroup and $\BZ_2$ 1-form symmetry, respectively. Since the $\HF$ data are independent of the choice of $(W,\widetilde{W})$, we may fix $(W,\widetilde{W})=(1,1)$, thereby quotienting the solutions by the Weyl subgroup. This yields \eqref{eq: n=2 GL HF}.

\subsubsection{Bethe vacua for $n\geq 3$}
For $n\geq 3$, the Bethe equations \eqref{eq: BE without ansatz} reduce to the matrix equation
\begin{align}
\label{eq: n geq 3 matrix eq}
    \mathrm{K}^{\pm} [\vec{k}]\cdot\vec{X}^{\vec{G}}\in \p i \BZ^n
\end{align}
where $X_1^{\vec{G}}\equiv X_1$, $X_{i\geq 2}^{\vec{G}}\equiv (-1)^{\sum_{j=1}^{i-1} G_j} X_i$ and
\begin{align} \label{eq: Global linear ansatz coefficient matrix}
    \mathrm{K}^{\pm}[\vec{k}]\equiv \left(\begin{array}{cccccccc}
        k_1&1&0&0&\cdots&0&0&\pm 1\\
        1&k_2&1&0&\cdots&0&0&0\\
        0&1&k_3&1&\cdots&0&0&0\\
        0&0&1&k_4&\cdots&0&0&0\\
        \vdots&\vdots&\vdots&\vdots&\ddots&\vdots&\vdots&\vdots\\
        0&0&0&0&\cdots&k_{n-2}&1&0\\
        0&0&0&0&\cdots&1&k_{n-1}&1\\
        \pm 1&0&0&0&\cdots&0&1&k_n
    \end{array}\right)\;,\quad\pm 1=(-1)^{\sum_{j=1}^n G_j}\;.
\end{align}
$\mathrm{K}^{\pm}[\vec{k}]$ satisfies $\det \mathrm{K}^{\pm}[\vec{k}]=\mathrm{tr} \vf\mp 2 (-1)^n$. For each $\vec{N}\in \p i \BZ^n$, one can verify that $\vec{X}_{\pm}^{\vec{G}}=\sum_l N_l \vec{V}_{\pm}^l$ is a solution to \eqref{eq: n geq 3 matrix eq}, where
\begin{align} \label{eq: Global linear for generic n}
\begin{split}
    \vec{V}_\pm^l& =\frac{\vec{q}^{\;2-l}}{\det{\rm K}^\pm[\vec{k}]}\text{ with }\\
    q_i&\equiv f^{\pm,l}_{i}\bigr[\pm(-1)^{i+n-1} (S T^{k_{l}}\cdots S T^{k_{l+i-2}})+(-1)^{i+1}(S T^{k_{l+i-1}}\cdots S T^{k_{l+n-1}})\bigr]_{12}\;,\\
    k_{i+n} & \equiv k_i\;,\qquad \; \; q_{i+n}\equiv q_i\;, \qquad \qquad \qquad q^l_i\equiv q_{l+i-1}\;,
    \\f_i^{+,l}&\equiv1\;, \quad \quad  \quad f_i^{-,l} \equiv\begin{cases}
        -1&(i<l)\;,
        \\1&(i\ge l)\;.
    \end{cases}
\end{split}
\end{align}
The associated $\HF$ data are given by
\begin{align} \label{eq: HF for generic n global}
    \CH_\pm=2^n\left|\det{\rm K}^\pm[\vec{k}]\right|\;,\quad\CF_\pm=\exp\left(\frac{5n\pi i}{12}+\frac{1}{2\pi i}\bigr(\vec{X}^{\vec{G}}_\pm\bigr)^T\cdot{\rm K^\pm}[\vec{k}]\cdot\vec{X}^{\vec{G}}_\pm\right)\;.
\end{align}
Since the $2^{n-1}$ choices of $\vec{G}$ within a fixed $(-1)^{\sum_{j=1}^n G_j}$ sector lead to the same equation for $\vec{X}^{\vec{G}}$, they give rise to $2^{n-1}$ distinct solutions (up to $2\pi i$ periodicity) that share the same physical $\HF$ datum. In addition, an extra $\{\pm1\}$-valued parameter $W$ can be introduced to label an overall sign flip of the solution. From \eqref{eq: HF for generic n global}, we find that the choice of $W$ also leaves the $\HF$ datum invariant\footnote{See $W$ in \eqref{eq: n=2 global linear Km} and \eqref{eq: n=2 global linear Kp}, for example.}. Then $2^n$ Weyl degrees of freedom shown in \eqref{eq: Weyl and Z2 1fs} can be labeled by
\begin{align} \begin{split}
    (W,G_1,G_n)\longmapsto(-W,1-G_1,1-G_n)&\text{ for SU}(2)^1\;,
    \\(G_{i-1},G_{i})\longmapsto(1-G_{i-1},1-G_{i})\phantom{jjj}&\text{ for SU}(2)^{i=2,3,\cdots,n}\;.
\end{split} \end{align}

\subsection{Local linear ansatz} \label{subsec: Local ansatz appendix}
We now turn to another type of ansatz, which we call the local linear ansatz. We assume that, near the A-twist point $\n=-1+\e$, the solutions take the form
\begin{align}
    Z_i=\sum_{n=0}^{\infty} Z_i^{(n)}\e^n\;,\quad X_i=\sum_{n=0}^{\infty} X_i^{(n)}\e^n
\end{align}
for $i=1,2,\cdots,n$. For a systematic analysis, we define
\begin{align}
    \D_i\equiv Z_i-X_i\equiv \sum_{n=0}^{\infty} \D_i^{(n)} \e^n\;,\quad \G_i\equiv Z_i+X_i\equiv \sum_{n=0}^{\infty} \G_i^{(n)} \e^n\;.
\end{align}
The Bethe equations \eqref{eq: BE without ansatz} yield four possibilities at order $\e^0$, depending on the values of $\D_i^{(0)}$ and $\G_i^{(0)}$:
\begin{itemize}
    \item Option A: For $\D_i^{(0)}, \G_i^{(0)}\notin 2\p i \BZ$, we find
    \begin{align}
        X_{i+1}^{(0)}\;,\quad k_iX_i^{(0)}+Z_{i-1}^{(0)}\in \pi i\mathbb{Z}\;.
    \end{align}
    \item Option B: For $\D_i^{(0)}\in 2\p i \BZ,\; \G_i^{(0)}\notin 2\p i \BZ$, we find
    \begin{align}
        2X_{i+1}^{(0)}+\log\left[\frac{2\D_i^{(1)}+\pi i}{-2\D_i^{(1)}+\pi i}\right]\;,\quad 2k_iX_i^{(0)}+2Z_{i-1}^{(0)}-\log\left[\frac{2\D_i^{(1)}+\pi i}{-2\D_i^{(1)}+\pi i}\right]\in \pi i(2\mathbb{Z}+1)\;.
    \end{align}
    \item Option C: For $\D_i^{(0)}\notin 2\p i \BZ,\; \G_i^{(0)}\in 2\p i \BZ$, we find
    \begin{align}
        2X_{i+1}^{(0)}+\log\left[\frac{2\G_i^{(1)}+\pi i}{-2\G_i^{(1)}+\pi i}\right]\;,\quad 2k_iX_i^{(0)}+2Z_{i-1}^{(0)}+\log\left[\frac{2\G_i^{(1)}+\pi i}{-2\G_i^{(1)}+\pi i}\right]\in \pi i(2\mathbb{Z}+1)\;.
    \end{align}
    \item Option D: For $\D_i^{(0)}, \G_i^{(0)}\in 2\p i \BZ$, we find
    \begin{align}
        \begin{split}
            2X_{i+1}^{(0)}+\log\left[\frac{2\D_i^{(1)}+\pi i}{-2\D_i^{(1)}+\pi i}\right]+\log\left[\frac{2\G_i^{(1)}+\pi i}{-2\G_i^{(1)}+\pi i}\right]&\in 2\pi i\mathbb{Z}\;,
            \\ 2Z_{i-1}^{(0)}-\log\left[\frac{2\D_i^{(1)}+\pi i}{-2\D_i^{(1)}+\pi i}\right]+\log\left[\frac{2\G_i^{(1)}+\pi i}{-2\G_i^{(1)}+\pi i}\right]&\in 2\pi i\mathbb{Z}\;.
        \end{split}
    \end{align}
    We introduce a $\BZ_2$-valued parameter $H_i$ defined by
    \begin{align}
        Z_i^{(0)}=X_i^{(0)}=\p i H_i\;.
    \end{align}
\end{itemize}
Options B and C have already appeared in the global linear ansatz. In option B(C), no solution exists for $\D_i^{(1)}$($\G_i^{(0)}$) if $X_{i+1}^{(0)}\in \p i \BZ$ or $k_i X_i^{(0)}+Z_{i-1}^{(0)}\in \p i \BZ$. Hence, options B and C constrain the values of $X_{i+1}^{(0)}$ and $Z_{i-1}^{(0)}$. One can construct various local ansatze by choosing one of $\{A,B,C,D\}$ for each $(Z_i,X_i)$, provided that these choices are mutually consistent.

\subsubsection{Bethe vacua for $n=2$ with $(k_1,k_2)\in (2\BZ+1)^2$}
For $n=2$, in addition to the global linear ansatz, one may choose the local ansatze $(A,D)$ and $(D,A)$. For the choice $(A,D)$, expanding the Bethe equations \eqref{eq: BE without ansatz} up to order $\e^3$ yields $(|k_1|-1)\times 2^3$ independent solutions labeled by $\a_1\in \{1,2,\cdots,\frac{|k_1|-1}{2}\}$, $W, \widetilde{W}\in \{\pm1\}$ and $\vec{H}\in\BZ_2^2$:
\begin{align}
\begin{split}
    \vec{X}= 2\pi iW\left(\frac{\a_1}{k_1},0\right)+\pi i\vec{H}\;,& \quad  Z_1=\log\left[\frac{1+i\widetilde{W}\sinh X_1}{\cosh X_1}\right]\;,\quad Z_2=X_2\;,\\
    \CH=4 |k_1 k_2|\;, &\quad \; \CF=\exp\paren{\frac{5\p i}{6}+\frac{2\p i}{k_1}\a_1^2}\;.
\end{split}
\end{align}
For the choice $(D,A)$, expanding the Bethe equations \eqref{eq: BE without ansatz} up to order $\e^3$ yields $(|k_2|-1)\times 2^3$ independent solutions labeled by $\a_2\in \{1,2,\cdots,\frac{|k_2|-1}{2}\}$, $W, \widetilde{W}\in \{\pm1\}$ and $\vec{H}\in\BZ_2^2$:
\begin{align}
    \vec{X}= 2\pi iW\left(0,\frac{\a_2}{k_2}\right)+\pi i\vec{H}\;, & \quad Z_2=\log\left[\frac{1+i\widetilde{W}\sinh X_2}{\cosh X_2}\right]\;,\quad Z_1=X_1\;,\\
    \CH=4 |k_1 k_2|\;, &\quad \; \CF=\exp\paren{\frac{5\p i}{6}+\frac{2\p i}{k_2}\a_2^2}\;.
\end{align}
$(W,\widetilde{W})$ and $\vec{H}$ implement the action of Weyl subgroup and $\BZ_2$ 1-form symmetry, respectively. Since the $\HF$ data are independent of the choice of $(W,\widetilde{W})$, we may fix $(W,\widetilde{W})=(1,1)$, thereby quotienting the solutions by the Weyl subgroup. This yields \eqref{eq: n=2 LL HF}.

\subsection{Global square root ansatz}
We next consider the global square root ansatz. We assume that, near the A-twist point $\n=-1+\e$, the solutions take the form
\begin{align}
    Z_i=\pi iH_i+g_i^Z\e^{1/2}+\CO(\e)\;,\quad X_i=\pi iH_i+g_i^X\e^{1/2}+\CO(\e)\;,\quad H_i\in\mathbb{Z}_2
\end{align}
for $i=1,2,\cdots, n$. The Bethe equations \eqref{eq: BE without ansatz} reduce to
\begin{align}
    \begin{split}
        2\left[g_{i+1}^X+\frac{\pi ig_i^Z}{(g_i^Z)^2-(g_i^X)^2}\right]\e^{1/2}+\CO(\e)&\in2\pi i\mathbb{Z}\;,
        \\ 2\left[g^Z_{i-1}+k_ig_i^X+\frac{\pi ig_i^X}{(g_i^X)^2-(g_i^Z)^2}\right]\e^{1/2}+\CO(\e)&\in2\pi i\mathbb{Z}
    \end{split}
\end{align}
and we get $\CA_i=\CB_i=0$ where
\begin{align} \label{eq: Global sqrt equation}
    \CA_i\equiv g^X_{i+1}+\frac{\pi ig_i^Z}{(g_i^Z)^2-(g_i^X)^2}\;,\quad \CB_i\equiv g^Z_{i-1}+k_ig_i^X+\frac{\pi ig_i^Z}{(g_i^Z)^2-(g_i^X)^2}\;.
\end{align}
With the definition $t_i\equiv g_{i+1}^X/g_i^Z$, the above relation becomes the recurrence relation
\begin{align}
\label{eq: ti recurrence}
    \frac{g_i^X \CA_i+g_i^Z \CB_i}{g_i^Z g_i^X}=t_i+t_{i-1}^{-1}+k_i=0
\end{align}
with $t_n=g_1^X/g_n^Z$ and $t_0\equiv t_n$. Then, $t_n$ can be written as
\begin{align}
    \label{eq: Equation for determine tn}
    t_n=-\frac{1}{k_{n+1}+t_{n+1}=k_1+t_1}=\cdots=-\frac{1}{k_{1}-\frac{1}{k_{2}-\cdots\frac{1}{k_{n}+t_{n}}}}=\frac{\vf_{21}+\vf_{22}t_n}{\vf_{11}+\vf_{12}t_n}\;.
\end{align}
The quadratic equation for $t_n$ has two solutions labeled by $s\in \{\pm 1\}$. For each choice of $s$, the remaining $t_i$, for $i=1,2,\cdots, n-1$, are uniquely determined by the recurrence relation \eqref{eq: ti recurrence}. With all $t_i$ thus determined, the solutions to $\CA_i=\CB_i=0$ can be written as
\begin{align}
    g_i^X=W_i\sqrt{s'\frac{\pi i [\vf(\vec{k}^{i+1})]_{12}}{\sqrt{p_+p_-}}}\;,\quad g_i^Z=\frac{g_{i+1}^X}{t_i}
\end{align}
where $W_i\in \{\pm 1\}$, $\vf(\vec{k}^{i+1})=ST^{k_{i+1}}\cdots ST^{k_{i+n}=k_i}$. Then, there are $2^{2 n+1}$ independent solutions labeled by $s'\in \{\pm 1\}$, $\vec{W}\in \{\pm 1\}^{\otimes n}$, and $\vec{H}\in \BZ_2^n$. $\vec{W}$ and $\vec{H}$ implement the action of Weyl subgroup and $\BZ_2$ 1-form symmetry, respectively. The associated $\HF$ data are given by
\begin{align}
    \CH=2^{n-2} p_+ p_- \paren{|p_+|+|p_-|+2 s \sqrt{p_+ p_-}}\;,\quad \CF=\exp\paren{\frac{5n\p i}{12}+\frac{\p i}{2}\vec{H}^T\cdot \mathrm{K}^+[\vec{k}]\cdot\vec{H}}
\end{align}
where $s=s' \mathrm{sign}(p_{\pm})$. Since the $\HF$ data are independent of the choice of $\vec{W}$, we may fix $\vec{W}=(1,1,\cdots,1)$, thereby quotienting the solutions by the Weyl subgroup. This yields \eqref{eq: HF data of sqrt ansatz}.

\section{Partition function integral}
\label{sec: Partition function integral}
The 3-sphere partition function of the $\BS(\vec{k})|_A$ theory can be computed directly via an (ordinary) integral obtained by supersymmetric localization. It provides a consistency check for the result of section~\ref{subsec: S(k) BG analysis} where we compute the 3-sphere partition function from the $\HF$ data. It is convenient to begin the computation near the B-twist point $\n=1+\e$. The squashed 3-sphere partition function of the $T[{\rm SU}(2)]$ theory near the B-twist point is
\begin{align}
\begin{split}
    &Z^{S^3_b}_{T[{\rm SU}(2)]}[X_1,X_2;m=0,\n=1+\e]
    \\&=\int\frac{dZ}{\sqrt{2\pi \hb}}e^{\frac{Z^2+X_1^2+2ZX_2}{\hb}}\QD\biggr[-\e\bigr(\pi i+\frac{\hb}{2}\bigr)\biggr]\prod_{\e_Z,\e_X\in\{\pm 1\}}\QD\biggr[\e_Z Z+\e_X X_1+\frac{2+\e}{2}\bigr(\pi i+\frac{\hb}{2}\bigr)\biggr]\;.
\end{split}
\end{align}
Using quantum dilogarithm function identities \eqref{eq: QDL reflection} and \eqref{eq: QDL limit}, we can modify it as
\begin{align} \begin{split}
    \lim_{\e\ra 0}\e Z^{S^3_b}_{T[{\rm SU}(2)]}=&-b\bigr(\pi i+\frac{\hb}{2}\bigr)^{-1}e^{\frac{\pi i}{12}(3+b^2+b^{-2})}\int\frac{dZ}{\sqrt{2\pi \hb}}e^{\frac{2ZX_2}{\hb}}
    \\ &\times\left[1+\frac{\e}{2}\bigr(\pi i+\frac{\hb}{2}\bigr)\sum_{\e_Z,\e_X\in\{\pm1\}}\frac{\QD'\bigr(\e_Z Z+\e_X X_1+\pi i+\frac{\hb}{2}\bigr)}{\QD\bigr(\e_Z Z+\e_X X_1+\pi i+\frac{\hb}{2}\bigr)}+\CO(\e^2)\right]\;.
\end{split} \end{align}
Then, with the help of one more identity
\begin{align}
\begin{split}
    \int \frac{dY}{\sqrt{2\pi \hb}}e^{\frac{2YX_2}{\hb}}\frac{\QD'\bigr(Y+\pi i+\frac{\hb}{2}\bigr)}{\QD\bigr(Y+\pi i+\frac{\hb}{2}\bigr)}&=\int_{\mathbb{R}+i0+}\frac{idt}{4\pi b}\frac{1}{\sinh(bt)\sinh(b^{-1}t)}\int\frac{dY}{\sqrt{2\pi \hb}}e^{\bigr(\frac{2X_2}{\hb}+\frac{it}{\pi b}\bigr)Y}
    \\&=\frac{e^{\frac{\pi i}{4}}}{4b}\frac{1}{\sinh(X_2)\sinh(X_2/b^2)}\;,
\end{split}
\end{align}
we find
\begin{align} \label{eq: TSU2 at Btwist}
    \lim_{\e\ra 0} Z^{S^3_b}_{T[{\rm SU}(2)]}=\frac{\#_b}{\e}\d(X_2)+\frac{e^{\frac{\pi i}{12}(b^2+b^{-2})-\frac{\pi i}{2}}}{2\sinh(X_2)\sinh(X_2/b^2)}\cos\biggr(\frac{X_1X_2}{\pi b^2}\biggr)+\CO(\e)
\end{align}
where $\#_b$ is a constant that depends on $b$ but is irrelevant to the final result. We now use the mirror property of the $T[{\rm SU}(2)]$ partition function\cite{Gang:2021hrd} to convert \eqref{eq: TSU2 at Btwist} into the corresponding expression near the A-twist point $\n=-1+\e$
\begin{align} \label{eq: Mirror relation}
    \lim_{\e\ra 0}Z^{S^3_b}_{T[{\rm SU}(2)]}[X_1,X_2;m=0,\n=-1+\e]=\lim_{\e\ra 0}Z^{S^3_b}_{T[{\rm SU}(2)]}[X_2,X_1;m=0,\n=1-\e]\;.
\end{align}
Using \eqref{eq: Mirror relation}, we can evaluate the squashed 3-sphere partition function of the $\BS(\vec{k})$ theory at the A-twist point as
\begin{align} \label{eq: S3 ptf from integral}
\begin{split}
    &Z^{S^3_b}_{\BS(\vec{k})|_A}=e^{\frac{n\pi i}{12}(b^2+b^{-2})-\frac{n\pi i}{2}}\int\prod_{i=1}^n\left[\frac{dX_i}{\sqrt{2\pi \hb}}e^{\frac{k_iX_i}{\hb}}\cos\biggr(\frac{X_iX_{i+1}}{\pi b^2}\biggr)\right]
    \\&\longrightarrow\left|Z^{S^3_b}_{\BS(\vec{k})|_A}\right|=\frac{1}{2^{\frac{n}{2}+1}}\bigr(\frac{1}{\sqrt{|p_+|}}+\frac{1}{\sqrt{|p_-|}}\bigr)\;.
\end{split}
\end{align}

\section{Quantum dilogarithm function} \label{sec: QDL}
On the squashed 3-sphere background introduced in section~\ref{subsec: S(k) BG analysis}, the contribution of the tetrahedron theory, which is the theory of a single chiral multiplet with the background $\CN=2$ supersymmetric Chern-Simons term of level $-\frac{1}{2}$ associated with its flavor ${\rm U}(1)$ symmetry, is given by the quantum dilogarithm function\cite{Dimofte:2011ju,Faddeev:1993rs} $\QD(Z)$ defined as
\begin{align}
    \QD(Z)\equiv\left\{\begin{matrix}
        \displaystyle\prod_{r=1}^\infty \frac{1-q^r e^{-Z}}{1-\Tilde{q}^{-r+1} e^{-\Tilde{Z}}} & \text{ if } & |q|<1\\
        \displaystyle\prod_{r=1}^\infty \frac{1-\Tilde{q}^r e^{-\Tilde{Z}}}{1-q^{-r+1} e^{-Z}} & \text{ if } & |q|>1
    \end{matrix}\right.\quad (\hb\equiv 2\pi ib^2)
\end{align}
with $q\equiv e^{2\pi ib^2}$, $\Tilde{q}\equiv e^{2\pi ib^{-2}}$, $\Tilde{Z}\equiv b^{-2}Z$. $Z$ is the real mass parameter associated with the background vector multiplet, rescaled by $2\pi b$. It satisfies
\begin{align}
    \QD(Z) \QD(-Z)=\frac{\exp\paren{\frac{i Z^2}{4\p b^2}-\frac{\p i}{6} (3+b^2+b^{-2})}}{4 \sinh\paren{\frac{Z}{2}} \paren{\frac{Z}{2 b^2}}}\;.
    \label{eq: QDL reflection}
\end{align}

\paragraph{Integral expression} The quantum dilogarithm function can be expressed as
\begin{align}
    \log\QD(Z)=\int_{\BR+i 0^+}  \frac{dt}{4t}\frac{\exp\paren{\frac{i t Z}{\p b}+t (b+b^{-1})}}{\sinh(bt) \sinh(b^{-1} t)}\;,\quad \textrm{ for }0<\Im[Z]<2\p (1+b^2)\;.
    \label{eq: QDL integral}
\end{align}
Using the difference equations
\begin{align}
    \QD(Z+2\p i b^2)=(1-e^{-Z})\QD(Z)\;,\quad \QD(Z+2\p i)=(1-e^{-Z/b^2})\QD(Z)\;,
\end{align}
we can compute all values of $\QD(Z)$ for finite real $b$.

\paragraph{Asymptotics} In the $b^2\ra0$ limit, where $Z\ra 0$, the quantum dilogarithm function satisfies
\begin{align}\label{eq: QDL limit}
        \lim_{Z\ra 0} Z\QD(Z) =  b\exp\paren{\frac{\p i}{4}-\frac{\p i}{12} (b^2+b^{-2})}\;.
\end{align}
For $\QD$ itself, the asymptotic expansion as $b\ra 0$ is given by
\begin{align}
    \log \QD(Z)\lora \sum_{n=0}^\infty \frac{B_n \hb^{n-1}}{n!}\Li_{2-n} \paren{e^{-Z}}\;.
\end{align}
Here $B_n$ is the $n$-th Bernoulli number with $B_1=\frac{1}{2}$.

\bibliographystyle{JHEP}
\bibliography{biblio.bib}

\end{document}